\RequirePackage{fix-cm}
\documentclass[smallextended]{svjour3}       
\smartqed  
\usepackage{graphicx}
\usepackage{subfigure}
\usepackage{amssymb}
\usepackage{color}
\usepackage{float}
\usepackage{ifpdf}
\usepackage[colorlinks=true]{hyperref}
\usepackage{braket}

\begin{document}

\title{Quantum filtering for a two-level atom driven by two counter-propagating photons}

\titlerunning{Quantum filtering for a two-level atom...}        

\author{Zhiyuan Dong \and Guofeng Zhang \and\\ Nina H. Amini}


\institute{Zhiyuan Dong \and Guofeng Zhang \at Department of Applied Mathematics, The Hong Kong Polytechnic University, Hong Kong, China\\\email{Guofeng.Zhang@polyu.edu.hk}
           \and
           Nina H. Amini \at Laboratoire des signaux et syst\`{e}mes (L2S), CNRS-CentraleSup\'{e}lec-Universit\'{e} Paris-Sud, Universit\'{e} Paris-Saclay, 3, rue Joliot Curie, 91190 Gif-sur-Yvette, France}

\date{Received: date / Accepted: date}

\maketitle

\begin{abstract}
The purpose of this paper is to propose quantum filters for a two-level atom driven by two continuous-mode counter-propagating photons and under continuous measurements. Two scenarios of multiple measurements, 1) homodyne detection plus photodetection, and 2) two homodyne detections, are discussed. Filtering equations for both cases are derived explicitly. As demonstration, the two input photons with rising exponential and Gaussian pulse shapes are used to excite a two-level atom under two homodyne detection measurements. Simulations reveal scaling relations between atom-photon coupling and photonic pulse shape for maximum atomic excitation.

\end{abstract}

\section{Introduction}\label{intro}

Quantum filters, pioneered by Belavkin \cite{belavkin1989nondemolition,belavkin1992quantum,belavkin1995quantum,Belavkin13}, has attracted lots of attention over the past decades \cite{gardiner2004quantum,bouten2007introduction,barchielli2009quantum,wiseman2009quantum}. In quantum optics, quantum filters, also known as quantum trajectories or stochastic master equations, are of great importance in measurement feedback control \cite{wiseman2009quantum,ZJ12,GB13,ZLW+17}. The problems of quantum filtering for systems driven by Gaussian states, including vacuum state, thermal state, coherent state, and squeezed state, have been well studied, see, e.g., \cite{DPZG92,chia2011quantum,Nur14,woolley2015quantum}. On the other hand, as single- and multi-photon states can nowadays be generated in real experiments \cite{raymer2005pure,wang2011efficient,YKT+13,lodahl2015interfacing,GZ15,OOM+16,lodahl2017chiral,sun2018single}, more and more research has been concentrated on atomic excitation \cite{wang2011efficient,baragiola2012n,RS16,RG17} and quantum filter design \cite{GOUGH12QUANTUM,GJN13,SONG13MULTI,DZA2016b,gao2016fault,baragiola17quantum} for systems driven by single or multiple photons.

Single-photon states play a fundamental role in quantum information, quantum computation, quantum measurements and control \cite{loudon00,lodahl2015interfacing,lodahl2017chiral,sun2018single}. Quantum filtering for systems driven by fields in single-photon states has been investigated in \cite{GOUGH12QUANTUM,GJN12,GJN13,carvalho2012cavity,gao2016faulta,gao2016fault,gao2017an,DZA2018}, among others. Particularly, quantum filters have been used to analyze conditional phase shifts on an optical cavity in \cite{carvalho2012cavity}. The master equations and filter equations for single-photon quantum filtering have been derived explicitly by using Markovian or non-Markovian embeddings in \cite{GOUGH12QUANTUM,GJN12,GJN13}. Recently, imperfect measurements and vacuum noise have also been considered in single-photon quantum filter design \cite{DZA2016b,gao2016fault,DZA2018}.

In addition to single-photon input states, the problem of quantum filtering for systems driven by multi-photon states has been discussed in \cite{SONG13MULTI,GJN14,DZA2016a}. A general multi-photon filtering framework has been proposed in \cite{SONG13MULTI} for both homodyne detection measurement and photon-counting measurement. Moreover, multiple measurements have been used to overcome the undesired vacuum noise and improve estimation performance \cite{DZA2016a}. In this paper, we derive quantum filters for a two-level atom driven by two counter-propagating input photons. Two cases of combined  measurements have been considered, namely 1) joint homodyne and  photon-counting measurements and 2) two homodyne measurements. The explicit forms of the quantum filters are given for both cases. As demonstration, the excitation probabilities have been simulated for a two-level atom which is driven by two counter-propagating photons with rising exponential and Gaussian pulse shapes, respectively. Several system parameters have been compared to achieve the optimal excitation probabilities in each scenario. For the single-photon filtering problem, it is well known that the optimal ratio for atomic excitation with a rising exponential pulse shape incident photon is $\gamma=\kappa$, where $\gamma$ is the full width at half maximum (FWHM) of the  photon wave packet and $\kappa$ is the decay rate  of the atom  \cite{SAL09,wang2011efficient,pan2016analysis}. On the other hand, if the incident photon is with Gaussian pulse shape, the optimal ratio is $\Omega=1.46\kappa$, where $\Omega$ is the photon bandwidth \cite{SAL09,RSF10,wang2011efficient,GOUGH12QUANTUM,baragiola2012n}. In this paper, when the two-level atom is driven by two counter-propagating identical photons, it is shown that the maximum of excitation probability attains at $\gamma=5\kappa$ for rising exponential pulse shapes (see the blue curve in Fig. \ref{rising}(a)), while $\Omega=2*1.46\kappa$ for the Gaussian pulse shapes (see Fig. \ref{Gau}).

Many features in the few-photon-atom interaction can be observed and analyzed in a simple way, such as master equations and their variants, see, e.g. \cite{RS16,RG17}. On the other hand, quantum filters  reveal the conditioned dynamics of the system under continuous measurements, which cannot be observed in a master equation, as the latter is the ensemble average of the former. Indeed, in section
\ref{sec:Gaussian},  numerical simulations have been performed for the study of a two-level atom driven by two counter-propagating photons with Gaussian pulse shapes. In each of the subfigures of Fig. \ref{Gau}, the fluctuating curves are individual quantum trajectories, whose average is given by the red solid curve in the top-right corner. Also, the master equation has been plotted in the black solid curve  in the top-right corner. It can be observed that in each subfigure the red solid curve and black solid curve almost coincide. This validates the fact that the master equation is the ensemble average of quantum trajectories, see Remark \ref{rem:master} in section
\ref{HH}. However, the quantum trajectories reveal more details of the dynamics that cannot be displayed by the master equation:
\begin{itemize}
\item  Some trajectories in Fig. \ref{Gau}(a) can reach almost unity probability, which means the two-level atom is fully excited in these scenarios.

\item  In Fig. \ref{Gau}(a),  the master equation (the black solid curve in the top-right corner) has the peak value at time $t=4$. The average of the quantum trajectories, given by the red solid curve  in the top-right corner, also reaches its maximum value at $t=4$. Nevertheless, many trajectories reach their maximum values at different time instants, although not far away from their mean value $t=4$. Indeed, this shows the stochastic nature of the atom-photon interaction.
\end{itemize}

Therefore, it is reasonable to say that a quantum filter is more powerful than a master equation, and is able to show more dynamics of an open quantum system under continuous measurements. Finally,  since quantum filters describe the conditional evolution of physical systems under continuous measurements, they are essential for real-time measurement-based feedback control of quantum systems \cite{wiseman2009quantum,amini2013feedback}.

This paper is organized as follows. In section \ref{preliminary}, we review some basic preliminaries such as the $(S,L,H)$ formalism, quantum filtering, and continuous-mode single-photon states. System augmentation used in this paper is introduced in section \ref{scenario} and the result of quantum filtering for systems driven by vacuum input states \cite{woolley2015quantum} is briefly recalled in section \ref{vacuum}. Then filtering equations with multiple measurements for a two-level atom driven by two counter-propagating single-photons are derived explicitly. Specifically, the joint homodyne-photoncounting measurement case is discussed in section \ref{HP}, while the joint homodyne-homodyne measurement case is discussed in section \ref{HH}. The excitation probabilities of a two-level atom interacting with photons with rising exponential pulse shapes and Gaussian pulse shapes are numerically simulated and discussed in section \ref{simulation}. Section \ref{conclusion} concludes this paper.

\emph{Notation.} Let $|\eta\rangle$ be the initial state of the two-level atom and $|0\rangle$ be the vacuum field state. Given a column vector of operators or complex numbers $X=[x_1,\cdots,x_n]^T$, the adjoint operator or complex conjugate of $X$ is denoted by $X^\#=[x_1^\ast,\cdots,x_n^\ast]^T$, and $X^\dagger=(X^\#)^T$. The commutator between operators $A$ and $B$ is defined to be $[A,B]=AB-BA$. Two superoperators are
\begin{eqnarray*}
&\mathrm{Lindbladian}&:\mathcal{L}_GX\equiv-i[X,H]+\mathcal{D}_LX,\\
&\mathrm{Liouvillian}&:\mathcal{L}_G^\star\rho\equiv-i[H,\rho]+\mathcal{D}_L^\star\rho,
\end{eqnarray*}
where $\mathcal{D}_LX=L^\dagger XL-\frac{1}{2}(L^\dagger LX+XL^\dagger L)$, and $\mathcal{D}_L^\star\rho=L\rho L^\dagger-\frac{1}{2}(L^\dagger L\rho+\rho L^\dagger L)$. Finally, $\delta_{jk}$ is the Kronecker delta function and $\delta(t-r)$ is the Dirac delta function.

\section{Preliminary}\label{preliminary}

\subsection{Open quantum systems  in the $(S,L,H)$ formalism}  \label{subsec:network}

In this paper, the system under study is a two-level atom that is driven by two counter-propagating photons. This is an open quantum system for which the so-called $(S,L,H)$ formalism has proven very convenient \cite{gough:2009,ZJ12,TNP+12,SHT+13,CKS+17}. Here, $S$ is a scattering operator satisfying $S^\dagger S=SS^\dagger=I$, $L$ denotes the coupling between the system and field, and the initial system Hamiltonian is described by the self-adjoint operator $H$.

\begin{figure}[htb]
\centering
\includegraphics[width=0.4\textwidth]{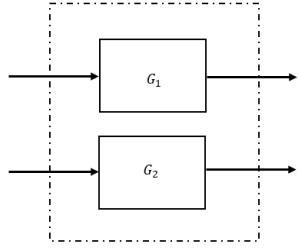}    
\caption{Concatenation product of two quantum systems $G_1$ and $G_2$}
\label{concate}
\end{figure}

Given two quantum systems $G_1=(S_1,L_1,H_1)$ and $G_2=(S_2,L_2,H_2)$, their concatenation product  $G_1\boxplus G_2$, as shown in Fig. \ref{concate}, is introduced in \cite{gough:2009}
\begin{equation}\nonumber
G_1\boxplus G_2=\left(\left[
                         \begin{array}{cc}
                           S_1 & 0 \\
                           0 & S_2 \\
                         \end{array}
                       \right]
,\left[
                          \begin{array}{c}
                            L_1 \\
                            L_2 \\
                          \end{array}
                        \right]
,H_1+H_2\right).
\end{equation}

\begin{figure}[htb]
\centering
\includegraphics[width=0.6\textwidth]{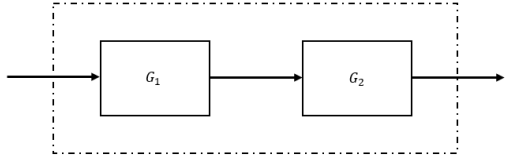}
\caption{Series product of two quantum systems $G_1$ and $G_2$}
\label{series}
\end{figure}

Moreover, if $G_1$ and $G_2$ have the same number of field channels, their series product $G_2\vartriangleleft G_1$, as shown in Fig. \ref{series}, can be defined by
\begin{equation}\nonumber
G_2\vartriangleleft G_1=\left(S_2S_1,L_2+S_2L_1,H_1+H_2+\mathrm{Im}\{L^\dag_2S_2L_1\}\right).
\end{equation}

\subsection{Quantum filtering} \label{subsec:filer}

As we study in this paper the problem of quantum filtering of a two-level atom driven by two counter-propagating photons, in this section we briefly introduce the basics of quantum filtering.

Let the annihilation operator for the $j$-th input field be $b_j(t)$ and its adjoint operator be $b_j^\ast(t)$. Since the field we consider is in continuous-mode, the following commutation relation holds:
\begin{equation} \label{eq:SCR}
\left[b_j(t),b_k^\ast(r)\right]=\delta_{jk}\delta(t-r), \ \ \ j,k=1,2.
\end{equation}
Denote $b(t) = \left[\begin{array}{c} b_1(t)\\ b_2(t)  \end{array}\right]$.
The integrated annihilation, creation, and gauge processes are given by
\[
B(t)=\int_{t_0}^tb(s)ds, ~ B^\#(t)=\int_{t_0}^tb^\#(s)ds, ~ \Lambda(t)=\int_{t_0}^tb^\#(s)b^T(s)ds,
\]
respectively, where $t_0$ is the time when the system and field start interaction. Due to (\ref{eq:SCR}), these  quantum stochastic processes satisfy
\begin{eqnarray}\nonumber
&&dB_j(t)dB_k^\ast(t)=\delta_{jk}dt,~~dB_j(t)d\Lambda_{kl}(t)=\delta_{jk}dB_l(t),\\\label{table}
&&d\Lambda_{jk}(t)dB_l^\ast(t)=\delta_{kl}dB_j^\ast(t),~~d\Lambda_{jk}(t)d\Lambda_{lm}(t)=\delta_{kl}d\Lambda_{jm}(t).
\end{eqnarray}

The unitary operator $U(t)$ on the tensor product Hilbert spaces $\mathrm{System}\otimes\mathrm{Field}$ can be used to describe the dynamical evolution of a  quantum system in the $(S,L,H)$ formalism, which is the solution to the quantum stochastic differential equation (QSDE)
\begin{equation}\label{QSDE}
dU(t)=\Bigg\{-\left(iH+\frac{1}{2}L^\dagger L\right)dt+LdB^\dagger(t)-L^\dagger SdB(t)+{\rm Tr}[S-I]d\Lambda(t)\Bigg\}U(t)
\end{equation}
with the initial condition $U(t_0)=I$ (identity operator).

Based on (\ref{table}) and (\ref{QSDE}), the time evolution of the system operator $X$, denoted by
\begin{equation}\label{eq:X}
j_t(X)\equiv X(t)=U^\dagger(t)(X_{\mathrm{system}}\otimes I_{\mathrm{field}})U(t),
\end{equation}
is given by
\begin{eqnarray}\nonumber
dj_t(X)&=&j_t(\mathcal{L}_GX)dt+dB^\dagger(t)j_t(S^\dagger[X,L])\\
&&+j_t([L^\dagger,X]S)dB(t)+{\rm Tr}[j_t(S^\dagger XS-X)d\Lambda(t)].
\end{eqnarray}
The dynamical evolution of the output field  in the input-output formalism is given by
\begin{eqnarray*}
dB_{\mathrm{out}}(t)&=&L(t)dt+S(t)dB(t),\\
d\Lambda_{\mathrm{out}}(t)&=&L^\#(t)L^T(t)dt+S^\#(t)dB^\#(t)L^T(t)\\
&&+L^\#(t)dB^T(t)S^T(t)+S^\#(t)d\Lambda(t)S^T(t),
\end{eqnarray*}
where $B_{\mathrm{out}}(t)=U^\dagger(t)(I_{\mathrm{system}}\otimes B(t))U(t)$ is the integrated output annihilation operator, and  $\Lambda_{\mathrm{out}}(t)=U^\dagger(t)(I_{\mathrm{system}}\otimes \Lambda(t))U(t)$ is the output gauge process.

The output fields can be continuously measured,  and from these measurements one can construct quantum filters to study the conditioned dynamics of the system. Homodyne detection and photon-counting measurements are commonly used in quantum optical experiments. In the case of homodyne detection, the measurement equation is
\begin{equation}
Y(t)=U^\dagger(t)(I_{\mathrm{system}}\otimes(B(t)+B^\#(t)))U(t),
\end{equation}
while in the case of photon-counting measurement
\begin{equation}
Y(t)=U^\dagger(t)(I_{\mathrm{system}}\otimes\Lambda(t))U(t).
\end{equation}
Both of them enjoy the self-non-demolition property
\begin{equation}
[Y(t),Y(r)]=0,~~t_0\leq r\leq t,
\end{equation}
and the  non-demolition property
\begin{equation}
[X(t),Y(s)]=0,~~t_0\leq s\leq t.
\end{equation}
The quantum conditional expectation is defined as
\begin{equation}
\hat{X}(t)\equiv\pi_t(X)=\mathbb{E}[j_t(X)|\mathcal{Y}_t],
\end{equation}
where $\mathbb{E}$ denotes the expectation and the commutative von Neumann algebra $\mathcal{Y}_t$ is generated by the past measurement observation $\{Y(s):t_0\leq s\leq t\}$. Simply speaking, the quantum filtering problem is to find the minimum of the least mean squares estimation $\mathbb{E}[\{\pi_t(X)-j_t(X)\}^2]$ for the system observables $j_t(X)$.  The conditioned system density operator $\rho(t)$ can be obtained by means of $\pi_t(X)=\mathrm{Tr}\left\{(\rho(t))^\dagger X\right\}$.  It turns out that $\rho(t)$ is a solution to a system of stochastic differential equations, which is called {\it quantum filter} in the quantum control community or {\it quantum trajectories} in the quantum optics community, see, e.g.,
\cite{belavkin1989nondemolition,bouten2007introduction,barchielli2009quantum,GJN13,SONG13MULTI,wiseman2009quantum,CKS+17}.

\subsection{Continuous-mode single-photon states}

Let $b^\ast(t)$ be the creation operator of a travelling light field.  Define an integrated creation operator $B^\ast(\xi)$ for a single photon with pulse shape $\xi(t)$ to be
\begin{equation}
B^\ast(\xi)=\int_{-\infty}^\infty\xi(t)b^\ast(t)dt,
\end{equation}
where $|\xi(t)|^2 dt$ is the probability of finding the photon in the time interval $[t,t+dt)$, which satisfies the normalization condition $\int_{-\infty}^\infty|\xi(t)|^2dt=1$. A continuous-mode single-photon state can be given by
\begin{equation}
|1_\xi\rangle=B^\ast(\xi)|0\rangle,
\end{equation}
which means that the operator $B^\ast(\xi)$ acts on the vacuum state $|0\rangle$ to generate a single-photon state for the travelling light field. It can be verified that
\begin{equation}
dB(t)|1_\xi\rangle=\xi(t)dt|0\rangle,~~d\Lambda(t)|1_\xi\rangle=\xi(t)dB^\ast(t)|0\rangle.
\end{equation}

The problem of quantum filtering for systems driven continuous-mode single-photon fields has been investigated in a series of papers, e.g., \cite{GOUGH12QUANTUM,GJN12,GJN13,carvalho2012cavity,gao2016faulta,gao2016fault,DZA2018}.

\section{Quantum Filtering with Multiple Measurements}\label{Main}

\begin{figure}
\centering
\includegraphics[width=0.8\textwidth]{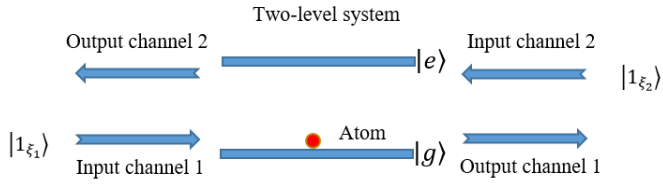}
\caption{A two-level atom is driven by two counter-propagating photons, one in each input channel. After interaction, the two photons are in the two output channels. Then the light fields in the two output channels are mixed by a beam splitter, the resulting light fields are measured. (The beam splitter and the measurement devices are not shown in this figure, but they can be found in Fig. \ref{depiction}.)}
\label{system}
\end{figure}

In this section, the filtering equations for a two-level atom driven by two counter-propagating continuous-mode photons and  under multiple measurements are derived explicitly, see Figs. \ref{system}-\ref{depiction}. The main ideas of the derivation of the quantum filters are summarized as follows. As shown in Fig. \ref{depiction}, two ancillas, $A_1$ and $A_2$, are used to generate single-photon states $|1_{\xi_1}\rangle$ and $|1_{\xi_2}\rangle$ from the vacuum input fields 1 and 2, the mechanism of such photon generation is discussed in details in section \ref{scenario}. The two photons generated interact with the two-level atom, denoted $G$ in Fig. \ref{depiction}, after interaction they are scattered into two output channels. These output channels are mixed by  a beam splitter to produce two output light fields (namely Output 1 and Output 2 in Fig. \ref{depiction}), which in the sequel are continuously measured by homodyne detectors or photodetectors. The augmented system $G_E$ (ancillas plus two-level atom plus beam splitter) in Fig. \ref{depiction} is driven by two vacuum input fields, the filter for this augmented system is presented in section \ref{vacuum}. However, as we are interested in the quantum filters for the two-level system under continuous measurements, we have to perform partial trace over the two ancillas $A_1$ and $A_2$, which is done in sections \ref{HP} and \ref{HH}. More specifically, in section \ref{HP}, the quantum filter is derived when the output fields are measured by a homodyne detector and  a photodetector, while in section \ref{HH}, the quantum filter is derived when the output fields are measured by two homodyne detectors.

\subsection{System augmentation}\label{scenario}

\begin{figure}
\includegraphics[width=0.9\textwidth]{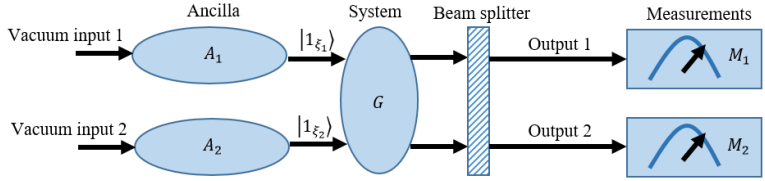}
\caption{Quantum system depiction: $A_1$, $A_2$ are the two ancillas which cascade the quantum system $G$, two measurements $M_1$ and $M_2$ will be used as the detectors, e.g., homodyne detector or photodetector. The augmented system (from the vacuum input to measurement) is denoted by $G_E$ for later use.}
\label{depiction}
\end{figure}

Fig. \ref{depiction} presents a more detailed quantum system depiction for the scheme shown in Fig. \ref{system}. Two quantum signal generators (ancillas $A_1$, $A_2$) are used to cascade the two-level atom $G$. The augmented system $G_E=\mathbf{S}_b\vartriangleleft G\vartriangleleft(A_1\boxplus A_2)$ could be obtained by means of the concatenation and series products introduced in section \ref{subsec:network}.  In what follows, we specify the parameter for the augmented system $G_E$. First, consider the beam splitter  in Fig. \ref{depiction}, which in the $(S,L,H)$ formalism  is given by $\mathbf{S}_b=(S_b,0,0)$ with
\begin{equation}\label{beamsplitter}
S_b=\left[
             \begin{array}{cc}
               \sqrt{1-r^2} & r \\
               -r & \sqrt{1-r^2} \\
             \end{array}
           \right],~~0\leq r\leq1.
\end{equation}
Next, we consider the two-level atom $G$.  In this paper, we consider the {\it on-resonance} case, i.e., the carrier frequency of the single-photon light field is equal to the transition frequency between the excited and ground states of the two-level atom. As a result, the system Hamiltonian $H=0$, which maximizes the atom-photon interaction \cite{wang2011efficient,baragiola2012n,RG17}. In the $(S,L,H)$ formalism, we have $G=(I_2,L,0)$, where $L=\left[
   \begin{array}{c}
     \sqrt{\kappa_1} \\
     \sqrt{\kappa_2} \\
   \end{array}
 \right]\sigma_-
$, where $\sigma_-$ is the lowering operator of the two-level atom $G$. Finally, we look at the two ancillas $A_1$ and $A_2$, who in the $(SL,H)$ formalism are given by
\begin{equation}
A_1=(I,L_1,0), ~~A_2=(I,L_2,0),
\end{equation}
where the coupling operators are $L_1=\lambda_1(t)\sigma_{-1}$ and  $L_2=\lambda_2(t)\sigma_{-2}$, respectively. Here, $\sigma_{-1}$ and $\sigma_{-2}$ correspond to lowering operators from the excited state $\ket{\uparrow}$ to the ground state $\ket{\downarrow}$ for the two ancillas. The rising operators will be denoted by $\sigma_{+1}$ and $\sigma_{+2}$ respectively. Ordinary functions $\lambda_1(t)$ and  $\lambda_2(t)$ are given by
\begin{equation}
\lambda_1(t)=\frac{\xi_1(t)}{\sqrt{w_1(t)}},~~\lambda_2(t)=\frac{\xi_2(t)}{\sqrt{w_2(t)}},
\end{equation}
where $w_1(t)=\int_t^\infty|\xi_1(s)|^2ds$, $w_2(t)=\int_t^\infty|\xi_2(s)|^2ds$, and $\xi_1(t)$ and $\xi_2(t)$ are the corresponding input pulse shapes in the first and second channels. As shown in \cite{GOUGH12QUANTUM,GZ15,DZA2018},  if the ancilla $A_i$ $(i=1,2)$ is initialized in the excited state $\ket{\uparrow}$ and driven by a vacuum field, it will generate a single photon state $\ket{1_{\xi_i}}$ $(i=1,2)$. That is why $A_1$ and $A_2$ are called signal generators.

By the concatenation and series products introduced in section \ref{subsec:network}, the augmented system $G_E =\mathbf{S}_b\vartriangleleft G\vartriangleleft(A_1\boxplus A_2)$ in Fig. \ref{depiction} can be re-parameterized as
\begin{equation}
G_E =(S_t,L_t,H_t),
\end{equation}
where
\begin{eqnarray*}
S_t&=&\left[
      \begin{array}{cc}
        \sqrt{1-r^2} & r \\
        -r & \sqrt{1-r^2} \\
      \end{array}
    \right],\\
L_t&=&\left[
      \begin{array}{c}
        \sqrt{1-r^2}(L_1+\sqrt{\kappa_1}\sigma_-)+r(L_2+\sqrt{\kappa_2}\sigma_-) \\
        -r(L_1+\sqrt{\kappa_1}\sigma_-)+\sqrt{1-r^2}(L_2+\sqrt{\kappa_2}\sigma_-) \\
      \end{array}
    \right],\\
H_t&=&\frac{1}{2i}(\sqrt{\kappa_1}\sigma_+L_1+\sqrt{\kappa_2}\sigma_+L_2-\sqrt{\kappa_1}L_1^\dagger\sigma_--\sqrt{\kappa_2}L_2^\dagger\sigma_-).
\end{eqnarray*}

In the following, we use $\tilde{U}(t)$ to describe the evolution operator for the  augmented system $G_E$. Then the following equality can be verified
\begin{equation} \label{ex:X_extended}
\mathbb{E}_{\eta\xi}[X(t)]=\mathbb{E}_{\uparrow\eta0}[\tilde{U}^\dagger(t)(I_{\rm ancilla}\otimes X\otimes I_{\rm field})\tilde{U}(t)],
\end{equation}
where on the right hand side, $\uparrow$ means that the ancillas are initialized in the excited state, $\eta$ is the initial state of the two-level atom, and $0$ denotes the vacuum input field state.  In the next section, we present the quantum filter for the augmented system $G_E$.

\subsection{Quantum filter for the augmented system $G_E$}\label{vacuum}

By means of system augmentation introduced in the preceding section, the augmented system $G_E$ is now driven by two vacuum fields, see Fig. \ref{depiction}. The problem of quantum filtering for this type of systems has been studied in \cite{woolley2015quantum}. In this section, we cite the main result of \cite{woolley2015quantum} in order to derive quantum filters for a two-level atom driven by two counter-propagating photons in sections \ref{HP} and \ref{HH}.

As discussed in \cite{woolley2015quantum}, a general measurement equation, which is a function of annihilation, creation and gauge processes of the output fields, may be defined as
\begin{equation}\label{general}
dY(t)=F_1^\#dB_{\mathrm{out}}^\#(t)+F_1dB_{\mathrm{out}}(t)+F_2\mathrm{diag}(d\Lambda_{\mathrm{out}}(t)).
\end{equation}
Moreover, a set of measurements $Y(t)$ is self-commutative if and only if the constant matrices  $F_1$ and $F_2$ satisfy
\begin{eqnarray*}
\left[
  \begin{array}{cc}
    F_1 & F_1^\# \\
  \end{array}
\right]\left[
         \begin{array}{cc}
           \mathbf{0} & I \\
           -I & \mathbf{0} \\
         \end{array}
       \right]\left[
                \begin{array}{c}
                  F_1^T \\
                  F_1^\dagger \\
                \end{array}
              \right]&=&\mathbf{0}, \\
\left[
  \begin{array}{cc}
    F_2 & F_1^\# \\
  \end{array}
\right]\left[
         \begin{array}{cc}
           \mathbf{0} & I \\
           -I & \mathbf{0} \\
         \end{array}
       \right]\left[
                \begin{array}{c}
                  F_2^T \\
                  F_1^\dagger \\
                \end{array}
              \right]&=&\mathbf{0}, \\
\left[
  \begin{array}{cc}
    F_2 & F_1 \\
  \end{array}
\right]\left[
         \begin{array}{cc}
           \mathbf{0} & I \\
           -I & \mathbf{0} \\
         \end{array}
       \right]\left[
                \begin{array}{c}
                  F_2^T \\
                  F_1^T \\
                \end{array}
              \right]&=&\mathbf{0}.
\end{eqnarray*}

In the following, we cite the main result in \cite{woolley2015quantum}, which presents the quantum filter for an open quantum system driven by vacuum input fields.

\begin{lemma}\cite[Theorem 3.2]{woolley2015quantum}\label{lem:woolley}
Let $\{Y_{i,t},i=1,2,\ldots,N\}$ be a set of $N$ compatible measurement outputs for a quantum system $\mathcal{E}$, i.e., $Y_{i,t}$ satisfy the self-commutative and non-demolition properties. With {\bf vacuum} input state, the quantum filter is given by
\begin{equation}
d\tilde{X}_t=\pi_t[\mathcal{L_E}(\tilde{X}_t)]dt+\displaystyle{\sum^N_{i=1}}\beta_{i,t}dW_{i,t},
\end{equation}
where $dW_{i,t}=dY_{i,t}-\pi_t(dY_{i,t})$ is a martingale process for each measurement output and $\beta_{i,t}$ is the corresponding gain given by
\begin{eqnarray}
\zeta^T&=&\pi_t(\tilde{X}_tdY^T_t)-\pi_t(\tilde{X}_t)\pi_t(dY^T_t)+\pi_t\left([L^\dag_t,\tilde{X}_t]S_tdBdY^T_t\right),\\
\Sigma&=&\pi_t(dY_tdY^T_t), ~~ \beta=\Sigma^{-1}\zeta
\end{eqnarray}
with $\Sigma$ being assumed to be non-singular.
\end{lemma}

The augmented system $G_E$ discussed in section \ref{scenario} is an open quantum system driven by vacuum input fields. As a result, Lemma \ref{lem:woolley} can be used to derive its quantum filter. However, in this paper what we are interested in are quantum filters for a two-level atom driven by two continuous-mode counter-propagating photons. Hence, more developments have to be carried out, which are tasks for sections \ref{HP} and \ref{HH}.

\subsection{Quantum filter for the case of joint homodyne and photon-counting measurements} \label{HP}

In this section, we present the quantum filter for the case where the output fields are measured by a combination of homodyne detector and photodetector.

For the augmented system $G_E$ introduced in section \ref{scenario}, if we choose $F_1=\left[
                  \begin{array}{cc}
                    1 & 0 \\
                    0 & 0 \\
                  \end{array}
                \right]
$ and $F_2=\left[
           \begin{array}{cc}
             0 & 0 \\
             0 & 1 \\
           \end{array}
         \right]$
in the general measurement equation (\ref{general}), then we have
\begin{eqnarray}
dY_1(t)
&=&
\sqrt{1-r^2}\Big[dB_1(t)+dB_1^\ast(t)+(L_1+\sqrt{\kappa_1}\sigma_-)dt
\nonumber
\\
&&
+(L_1^\ast+\sqrt{\kappa_1}\sigma_+)dt\Big]+r\Big[dB_2(t)+dB_2^\ast(t)
\nonumber
\\
&&
+(L_2+\sqrt{\kappa_2}\sigma_-)dt+(L_2^\ast+\sqrt{\kappa_2}\sigma_+)dt\Big],
\label{HD11}
\\
dY_2(t)
&=&\Big[r^2(L_1^\ast+\sqrt{\kappa_1}\sigma_+)(L_1+\sqrt{\kappa_1}\sigma_-)
\nonumber
\\
&&
-r\sqrt{1-r^2}(L_2^\ast+\sqrt{\kappa_2}\sigma_+)(L_1+\sqrt{\kappa_1}\sigma_-)
\nonumber
\\
&&
-r\sqrt{1-r^2}(L_1^\ast+\sqrt{\kappa_1}\sigma_+)(L_2+\sqrt{\kappa_2}\sigma_-)
\nonumber
\\
&&
+(1-r^2)(L_2^\ast+\sqrt{\kappa_2}\sigma_+)(L_2+\sqrt{\kappa_2}\sigma_-)\Big]dt
\nonumber
\\
&&
+\big[r^2(L_1^\ast+\sqrt{\kappa_1}\sigma_+)-r\sqrt{1-r^2}(L_2^\ast+\sqrt{\kappa_2}\sigma_+)\big]dB_1(t)
\nonumber
\\
&&
+\big[r^2(L_1+\sqrt{\kappa_1}\sigma_-)-r\sqrt{1-r^2}(L_2+\sqrt{\kappa_2}\sigma_-)\big]dB_1^\ast(t)
\nonumber
\\
&&
+\big[(1-r^2)(L_2^\ast+\sqrt{\kappa_2}\sigma_+)-r\sqrt{1-r^2}(L_1^\ast+\sqrt{\kappa_1}\sigma_+)\big]dB_2(t)
\nonumber
\\
&&
+\big[(1-r^2)(L_2+\sqrt{\kappa_2}\sigma_-)-r\sqrt{1-r^2}(L_1+\sqrt{\kappa_1}\sigma_-)\big]dB_2^\ast(t)\nonumber
\\
&&
+r^2d\Lambda_{11}(t)-r\sqrt{1-r^2}d\Lambda_{21}(t)-r\sqrt{1-r^2}d\Lambda_{12}(t)
\nonumber
\\
&&+(1-r^2)d\Lambda_{22}(t),
\label{HD12}
\end{eqnarray}
That is, $Y_1(t)$ is a mixture of the filed quadratures, while $Y_2(t)$ is a mixture of the filed quadratures and gauge processes.

Let $dY(t) =  [dY_1(t) \ dY_2(t)]^T$. By (\ref{table}), it is easy to see that
\begin{equation} \label{eq:YY^T}
\tilde{\pi}_t\left[dY(t)dY^T(t)\right]=\left[
                             \begin{array}{cc}
                               dt & 0 \\
                               0 & \tilde{\pi}_t\left[dY_2(t)dY_2(t)\right] \\
                             \end{array}
                           \right],
\end{equation}
where
\begin{eqnarray*}
\tilde{\pi}_t\left[dY_2(t)dY_2(t)\right]&=&r^2\tilde{\pi}_t\big[(L_1^\dag+\sqrt{\kappa_1}\sigma_+)(L_1+\sqrt{\kappa_1}\sigma_-)\big]dt\\
&&-r\sqrt{1-r^2}\tilde{\pi}_t\big[(L_1^\dag+\sqrt{\kappa_1}\sigma_+)(L_2+\sqrt{\kappa_2}\sigma_-)\big]dt\\
&&-r\sqrt{1-r^2}\tilde{\pi}_t\big[(L_2^\dag+\sqrt{\kappa_2}\sigma_+)(L_1+\sqrt{\kappa_1}\sigma_-)\big]dt\\
&&+(1-r^2)\tilde{\pi}_t\big[(L_2^\dag+\sqrt{\kappa_2}\sigma_+)(L_2+\sqrt{\kappa_2}\sigma_-)\big]dt.
\end{eqnarray*}

The Lindblad superoperator for the augmented system in Fig. \ref{depiction} can be expressed as
\begin{eqnarray}\nonumber
&&\mathcal{L}_{L_t}(A_1\otimes A_2\otimes X)\\\nonumber
&=&\mathcal{D}_{L_1}(A_1)\otimes A_2\otimes X+A_1\otimes\mathcal{D}_{L_2}(A_2)\otimes X\\\nonumber
&&+(\kappa_1+\kappa_2)A_1\otimes A_2\otimes\mathcal{D}_{\sigma_-}(X)\\\nonumber
&&+\sqrt{\kappa_1}L_1^\dag A_1\otimes A_2\otimes[X,\sigma_-]+\sqrt{\kappa_1}A_1L_1\otimes A_2\otimes[\sigma_+,X]\\\label{Lind}
&&+\sqrt{\kappa_2}A_1\otimes L_2^\dag A_2\otimes[X,\sigma_-]+\sqrt{\kappa_2}A_1\otimes A_2L_2\otimes[\sigma_+,X].
\end{eqnarray}

In what follows, let $\pi_t(X)$ be the conditional expectation of the operator $X$ of the two-level atom $G$ and $\tilde{\pi}_t (A_1\otimes A_2\otimes X)$ be the conditional expectation of the operator $A_1\otimes A_2\otimes X$   for the extended system $G_E$. Then it can be shown that
\begin{equation}\label{nota}
\pi_t^{jk;mn}(X)=\frac{\tilde{\pi}_t(Q_1^{jk}\otimes Q_2^{mn}\otimes X)}{w_1^{jk}(t)w_2^{mn}(t)},~~j,k,m,n=0,1,
\end{equation}
where $Q_1^{jk}$, $Q_2^{mn}$ are the operators of ancillas $A_1$, $A_2$, which are given by
\begin{equation}\nonumber
\left[
           \begin{array}{cc}
             Q_i^{00} & Q_i^{01} \\
             Q_i^{10} & Q_i^{11} \\
           \end{array}
         \right]=\left[
                   \begin{array}{cc}
                     \sigma_{+i}\sigma_{-i} & \sigma_{+i} \\
                     \sigma_{-i} & I \\
                   \end{array}
                 \right],
\end{equation}
and
\begin{equation}\nonumber
\left[
           \begin{array}{cc}
             w_i^{00} & w_i^{01} \\
             w_i^{10} & w_i^{11} \\
           \end{array}
         \right]=\left[
                   \begin{array}{cc}
                     w_i(t) & \sqrt{w_i(t)} \\
                     \sqrt{w_i(t)} & 1 \\
                   \end{array}
                 \right],~~i=1,2
\end{equation}
with $w_i(t)$ being given in section \ref{scenario}.

Based on (\ref{nota}) and Lemma \ref{lem:woolley},  the stochastic differential equations for the $\pi_t^{jk;mn}(X)$ can be derived. Consequently, the stochastic differential equations for the evolution of conditioned system density matrix $\rho^{jk;mn}(t)$ can be derived by means of the equivalence between the Schr\"{o}dinger picture and the Heisenberg picture $\pi_t^{jk;mn}(X)=\mathrm{Tr}\{(\rho^{jk;mn}(t))^\dagger X\},~~j,k,m,n=0,1.$ The following theorem presents the filtering equation of $\rho^{jk;mn}(t)$ when the output fields are under joint homodyne detection and photon-counting measurements, as given in (\ref{HD11})-(\ref{HD12}). The filtering equation for $\rho^{11;11}(t)$ is given  here and the others can be found in Appendix A.

\begin{theorem}\label{theorem1}
In the case of joint homodyne detection  and photon-counting measurements, the quantum filter for the two-level atom $G$ driven by two counter-propagating single-photon input states $|1_{\xi_i}\rangle$, $i=1,2$, is given by a system of stochastic differential equations, composed of (\ref{HD+Ph}) below and those in Appendix A.
\scriptsize
\begin{eqnarray}\nonumber
d\rho^{11;11}(t)&=&\Big\{(\kappa_1+\kappa_2)\mathcal{D}_{\sigma_-}^\star\rho^{11;11}(t)+\sqrt{\kappa_1}\xi_1(t)[\rho^{01;11}(t),\sigma_+]+\sqrt{\kappa_1}\xi_1^\ast(t)[\sigma_-,\rho^{10;11}(t)]\\\nonumber
&&+\sqrt{\kappa_2}\xi_2(t)[\rho^{11;01}(t),\sigma_+]+\sqrt{\kappa_2}\xi_2^\ast(t)[\sigma_-,\rho^{11;10}(t)]\Big\}dt\\\nonumber
&&+\Big\{\sqrt{1-r^2}\left[\xi_1^\ast(t)\rho^{10;11}(t)+\xi_1(t)\rho^{01;11}(t)+\sqrt{\kappa_1}\rho^{11;11}(t)\sigma_++\sqrt{\kappa_1}\sigma_-\rho^{11;11}(t)\right]\\\nonumber
&&+r\left[\xi_2^\ast(t)\rho^{11;10}(t)+\xi_2(t)\rho^{11;01}(t)+\sqrt{\kappa_2}\rho^{11;11}(t)\sigma_++\sqrt{\kappa_2}\sigma_-\rho^{11;11}(t)\right]\\\nonumber
&&-\rho^{11;11}(t)\left[\sqrt{1-r^2}z_{11}(t)+rz_{12}(t)\right]\Big\}dW_1(t)\\\nonumber
&&+\bigg\{K_p^{-1}(t)\Big\{r^2\Big[2|\xi_1(t)|^2\rho^{00;11}(t)+2\sqrt{\kappa_1}\xi_1(t)\rho^{01;11}(t)\sigma_++\sqrt{\kappa_1}\xi_1^\ast(t)\sigma_-\rho^{10;11}(t)\\\nonumber
&&+\kappa_1\sigma_-\rho^{11;11}(t)\sigma_+\Big]-r\sqrt{1-r^2}\Big[2\xi_1^\ast(t)\xi_2(t)\rho^{10;01}(t)+\sqrt{\kappa_2}\xi_1^\ast(t)\sigma_-\rho^{10;11}(t)\\\nonumber
&&+2\sqrt{\kappa_1}\xi_2(t)\rho^{11;01}(t)\sigma_++2\xi_1(t)\xi_2^\ast(t)\rho^{01;10}(t)+\sqrt{\kappa_1}\xi_2^\ast(t)\sigma_-\rho^{11;10}(t)\\\nonumber
&&+2\sqrt{\kappa_2}\xi_1(t)\rho^{01;11}(t)\sigma_+\Big]+(1-r^2)\Big[2|\xi_2(t)|^2\rho^{11;00}(t)+\sqrt{\kappa_2}\xi_2^\ast(t)\sigma_-\rho^{11;10}(t)\\
&&+2\sqrt{\kappa_2}\xi_2(t)\rho^{11;01}(t)\sigma_++\kappa_2\sigma_-\rho^{11;11}(t)\sigma_+\Big]\Big\}-\rho^{11;11}(t)\bigg\}dN(t),\label{HD+Ph}
\end{eqnarray}
\normalsize
where
\begin{eqnarray}
\rho^{01;11}(t)&=&(\rho^{10;11}(t))^\dagger,~~\rho^{11;01}(t)=(\rho^{11;10}(t))^\dagger,\\
\rho^{10;01}(t)&=&(\rho^{01;10}(t))^\dagger,~~\rho^{01;01}(t)=(\rho^{10;10}(t))^\dagger,\\
\rho^{00;01}(t)&=&(\rho^{00;10}(t))^\dagger,~~\rho^{01;00}(t)=(\rho^{10;00}(t))^\dagger,
\end{eqnarray}
with the initial conditions
\begin{equation}\nonumber
\rho^{11;11}(t_0)=\rho^{11;00}(t_0)=\rho^{00;11}(t_0)=\rho^{00;00}(t_0)=|\eta\rangle\langle\eta|.
\end{equation}
The innovation processes $dW_1(t)$ and $dN(t)$ are given by
\begin{eqnarray}
dW_1(t)&=&dY_1(t)-\Big[\sqrt{1-r^2}k_{11}(t)+rk_{12}(t)\Big]dt,\\
dN(t)&=&dY_2(t)-K_p(t)dt,
\end{eqnarray}
respectively, where
\begin{eqnarray*}
k_{11}(t)&=&\xi_1^\ast(t)\mathrm{Tr}[\rho^{10;11}(t)]+\xi_1(t)\mathrm{Tr}[\rho^{01;11}(t)]+\sqrt{\kappa_1}\mathrm{Tr}[(\sigma_++\sigma_-)\rho^{11;11}(t)],\\
k_{12}(t)&=&\xi_2^\ast(t)\mathrm{Tr}[\rho^{11;10}(t)]+\xi_2(t)\mathrm{Tr}[\rho^{11;01}(t)]+\sqrt{\kappa_2}\mathrm{Tr}[(\sigma_++\sigma_-)\rho^{11;11}(t)],\\
K_p(t)&=&r^2\mathrm{Tr}\Big[(L_1^\dag+\sqrt{\kappa_1}\sigma_+)(L_1+\sqrt{\kappa_1}\sigma_-)\rho^{11;00}(t)\Big]\\
&&-r\sqrt{1-r^2}\mathrm{Tr}\Big[(L_2^\dag+\sqrt{\kappa_2}\sigma_+)(L_1+\sqrt{\kappa_1}\sigma_-)\rho^{10;01}(t)\Big]\\
&&-r\sqrt{1-r^2}\mathrm{Tr}\Big[(L_1^\dag+\sqrt{\kappa_1}\sigma_+)(L_2+\sqrt{\kappa_2}\sigma_-)\rho^{01;10}(t)\Big]\\
&&+(1-r^2)\mathrm{Tr}\Big[(L_2^\dag+\sqrt{\kappa_2}\sigma_+)(L_2+\sqrt{\kappa_2}\sigma_-)\rho^{00;11}(t)\Big].
\end{eqnarray*}
\end{theorem}

\begin{remark}\label{rem:rho11a}
It can be verified that $\rho^{11;11}$ in Theorem \ref{theorem1} is indeed self-adjoint.
\end{remark}

\subsection{Quantum filter for the case of two homodyne detection measurements}\label{HH}

In this section, we consider the case that the output fields are measured by two homodyne detectors, that is, we choose $F_1=I_2$ and $F_2=0$ in the general measurement equation (\ref{general}). In this case,  the output fields in Fig. \ref{depiction} are $dY_1(t)$ as already given in (\ref{HD11}) and
\begin{eqnarray}\nonumber
dY_2(t)
&=&
-r\Big[dB_1(t)+dB_1^\dag(t)+(L_1+\sqrt{\kappa_1}\sigma_-)dt
\nonumber
\\
&&
+(L_1^\dag+\sqrt{\kappa_1}\sigma_+)dt\Big]+\sqrt{1-r^2}\Big[dB_2(t)+dB_2^\dag(t)
\nonumber
\\
&&
+(L_2+\sqrt{\kappa_2}\sigma_-)dt+(L_2^\dag+\sqrt{\kappa_2}\sigma_+)dt\Big].
\label{HH11}
\end{eqnarray}
Let $dY(t) =  [dY_1(t) \ dY_2(t)]^T$. By (\ref{table}), it is easy to see that
\begin{equation}
\tilde{\pi}_t\left[dY(t)dY^T(t)\right]=\left[
                             \begin{array}{cc}
                               dt & 0 \\
                               0 & dt \\
                             \end{array}
                           \right].
\end{equation}

The Lindblad superoperator in this case has the same form as that given in (\ref{Lind}). With the notation (\ref{nota}), we have the following theorem which presents the filtering equations for the conditional system density matrix. Similar as in Theorem \ref{theorem1},  here we only present the filtering equation for $\rho^{11;11}(t)$ while put the others in Appendix B.

\begin{theorem}\label{theorem}
In the case of two homodyne detection measurements, the quantum filter for the two-level system $G$ driven by two counter-propagating single-photon input states $|1_{\xi_i}\rangle$, $i=1,2$, is given by a system of stochastic differential equations, composed of (\ref{stochastic}) below and those in Appendix B.
\scriptsize
\begin{eqnarray}\nonumber
d\rho^{11;11}(t)&=&\Big\{(\kappa_1+\kappa_2)\mathcal{D}_{\sigma_-}^\star\rho^{11;11}(t)+\sqrt{\kappa_1}\xi_1(t)[\rho^{01;11}(t),\sigma_+]+\sqrt{\kappa_1}\xi_1^\ast(t)[\sigma_-,\rho^{10;11}(t)]\\\nonumber
&&+\sqrt{\kappa_2}\xi_2(t)[\rho^{11;01}(t),\sigma_+]+\sqrt{\kappa_2}\xi_2^\ast(t)[\sigma_-,\rho^{11;10}(t)]\Big\}dt\\\nonumber
&&+\Big\{\sqrt{1-r^2}\left[\xi_1^\ast(t)\rho^{10;11}(t)+\xi_1(t)\rho^{01;11}(t)+\sqrt{\kappa_1}\rho^{11;11}(t)\sigma_++\sqrt{\kappa_1}\sigma_-\rho^{11;11}(t)\right]\\\nonumber
&&+r\left[\xi_2^\ast(t)\rho^{11;10}(t)+\xi_2(t)\rho^{11;01}(t)+\sqrt{\kappa_2}\rho^{11;11}(t)\sigma_++\sqrt{\kappa_2}\sigma_-\rho^{11;11}(t)\right]\\\nonumber
&&-\rho^{11;11}(t)\left[\sqrt{1-r^2}z_{11}(t)+rz_{12}(t)\right]\Big\}dW_1(t)\\\nonumber
&&+\Big\{-r\left[\xi_1^\ast(t)\rho^{10;11}(t)+\xi_1(t)\rho^{01;11}(t)+\sqrt{\kappa_1}\rho^{11;11}(t)\sigma_++\sqrt{\kappa_1}\sigma_-\rho^{11;11}(t)\right]\\\nonumber
&&+\sqrt{1-r^2}\left[\xi_2^\ast(t)\rho^{11;10}(t)+\xi_2(t)\rho^{11;01}(t)+\sqrt{\kappa_2}\rho^{11;11}(t)\sigma_++\sqrt{\kappa_2}\sigma_-\rho^{11;11}(t)\right]\\
&&-\rho^{11;11}(t)\left[-rz_{11}(t)+\sqrt{1-r^2}z_{12}(t)\right]\Big\}dW_2(t),\label{stochastic}
\end{eqnarray}
\normalsize
where
\begin{eqnarray}
\rho^{01;11}(t)&=&(\rho^{10;11}(t))^\dagger,~~\rho^{11;01}(t)=(\rho^{11;10}(t))^\dagger,\\
\rho^{10;01}(t)&=&(\rho^{01;10}(t))^\dagger,~~\rho^{01;01}(t)=(\rho^{10;10}(t))^\dagger,\\
\rho^{00;01}(t)&=&(\rho^{00;10}(t))^\dagger,~~\rho^{01;00}(t)=(\rho^{10;00}(t))^\dagger,
\end{eqnarray}
with the initial conditions
\begin{equation}\nonumber
\rho^{11;11}(t_0)=\rho^{11;00}(t_0)=\rho^{00;11}(t_0)=\rho^{00;00}(t_0)=|\eta\rangle\langle\eta|.
\end{equation}
The innovation processes $dW_1(t)$ and $dW_2(t)$ are given by
\begin{eqnarray}
dW_1(t)&=&dY_1(t)-\left[\sqrt{1-r^2}z_{11}(t)+rz_{12}(t)\right]dt,\\
dW_2(t)&=&dY_2(t)+\left[rz_{11}(t)-\sqrt{1-r^2}z_{12}(t)\right]dt,
\end{eqnarray}
respectively, where
\begin{eqnarray*}
z_{11}(t)&=&\xi_1^\ast(t)\mathrm{Tr}[\rho^{10;11}(t)]+\xi_1(t)\mathrm{Tr}[\rho^{01;11}(t)]+\sqrt{\kappa_1}\mathrm{Tr}[(\sigma_++\sigma_-)\rho^{11;11}(t)],\\
z_{12}(t)&=&\xi_2^\ast(t)\mathrm{Tr}[\rho^{11;10}(t)]+\xi_2(t)\mathrm{Tr}[\rho^{11;01}(t)]+\sqrt{\kappa_2}\mathrm{Tr}[(\sigma_++\sigma_-)\rho^{11;11}(t)].
\end{eqnarray*}
\end{theorem}

\begin{remark}\label{rem:rho11b}
It can be verified that $\rho^{11;11}$ in Theorem \ref{theorem} is indeed self-adjoint.
\end{remark}

\begin{remark}\label{rem:master}
By partial tracing over the environment of the filtering equations in Theorems \ref{theorem1} and  \ref{theorem}, the terms involving innovation processes $dW_1(t)$, $dN(t)$ and $dW_2(t)$ will vanish. As a result, the filtering equations in Theorems \ref{theorem1} and \ref{theorem}   reduce to the same set of master equations. In other words, a quantum master equation is an ensemble average of a quantum filter.
\end{remark}

The following remark demonstrates that the scenario considered in this paper can be reduced to the case of a two-level atom driven by a single-channel single-photon state, which has been studied in \cite{GOUGH12QUANTUM,GJN12,GJN13,carvalho2012cavity,DZA2018}.

\begin{remark}\label{rem:1channel}
If we set $\kappa_1=0$ and the beam splitter $S_b=I$ in (\ref{beamsplitter}), then the filtering equations in Theorem \ref{theorem1} will reduce to the quantum filter for a two-level system driven by a single-photon state under photon-counting measurement, cf. \cite[Eq. (43)]{GOUGH12QUANTUM}, \cite[Eq. (3)]{carvalho2012cavity}. On the other hand, if we set $\kappa_2=0$ and the beam splitter $S_b=I$, then the filtering equations in both Theorems \ref{theorem1} and \ref{theorem} will reduce to quantum filters for a two-level system driven by a single-photon state under homodyne detection measurement, cf. \cite[Eq. (42)]{GOUGH12QUANTUM}, \cite[Eq. (2)]{carvalho2012cavity}.
\end{remark}

\section{Simulation Results}\label{simulation}

In this section, we employ the quantum filters derived above to calculate the excitation probabilities of a  two-level atom which is driven by two  counter-propagating photons and under two homodyne detection measurements. Assume that the two-level atom is initially in the ground state $|g\rangle$. Two types of photon pulse shapes, namely rising exponential and Gaussian pulse shapes, are considered.

\subsection{Rising exponential pulse shape}

The rising exponential pulse shapes of the two photons are given by
\begin{equation}\label{50}
\xi_i(t)=-\sqrt{\gamma_i}\ e^{\frac{\gamma_i}{2} t}H(-t),~~i=1,2,
\end{equation}
where $H(t)$ is the Heaviside function. These photons have Lorentzian lineshape functions with FWHM $\gamma_i$ $(i=1,2)$ \cite{bachor2004guide}.

\begin{figure}
\centering
\subfigure{\includegraphics[width=1.0\textwidth]{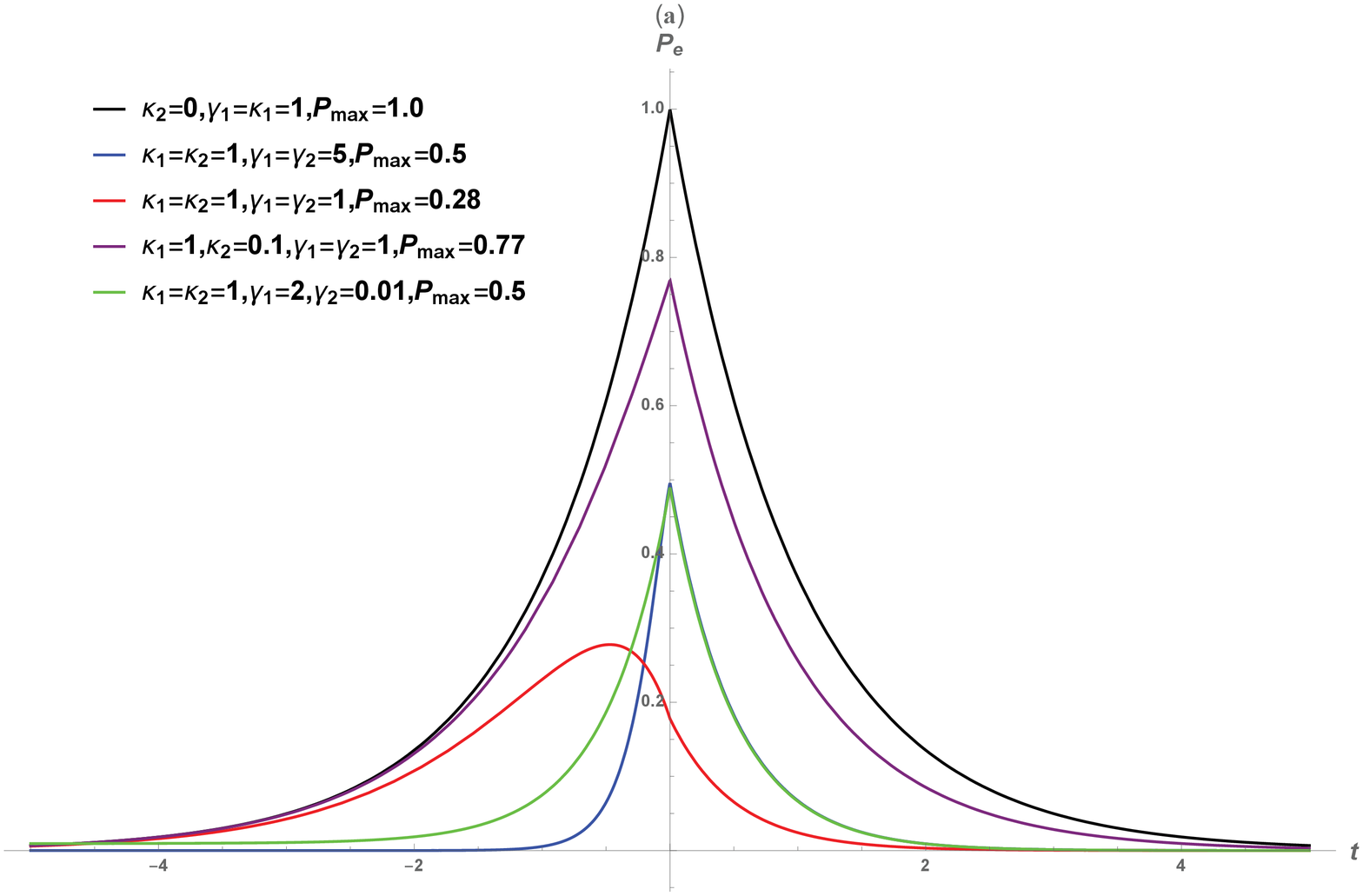}}
\vfill
\subfigure{\includegraphics[width=1.0\textwidth]{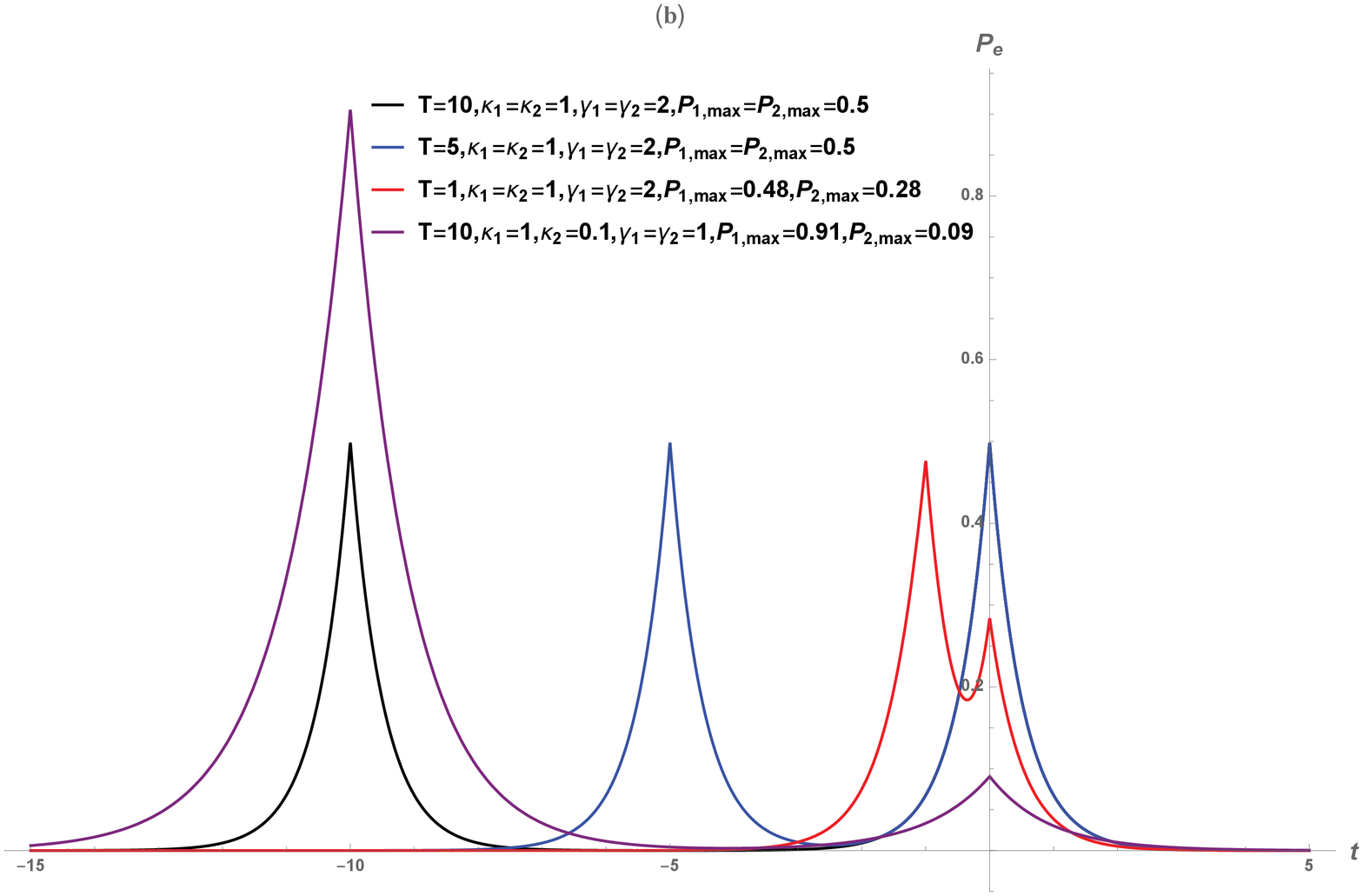}}
\caption{The excitation probabilities for a two-level atom driven by two counter-propagating photons  with rising exponential pulse shapes, calculated based on the master equation.}
\label{rising}
\end{figure}

In this example, we use the master equation to study  the excitation probabilities. As discussed in Remark \ref{rem:master}, the master equation can be obtained by tracing out the noise terms of the quantum filter in either Theorem \ref{theorem1} or Theorem \ref{theorem}. Excitation probabilities with various pulse shape parameters are plotted in Fig. \ref{rising}.  We have the following observations.
\begin{description}
\item[(i)] We look at a special case first. If we set $\kappa_2=0$, then, as discussed in  Remark \ref{rem:1channel}, we have the scenario that a two-level atom is driven by a single photon with a rising exponential pulse $\xi_1(t)$ defined in (\ref{50}). Let $\gamma_1=\kappa_1$. The excitation probability is given by the black curve in Fig. \ref{rising}(a). It can be seen that the excitation probability reaches its maximum $P_{\rm max}=1.0$ at time $t=0$, after that the excited atom decays to its ground state $|g\rangle$ exponentially. This is consistent with the well-known result that a two-level atom can be fully excited by an incident photon of a rising exponential pulse shape \cite{SAL09,wang2011efficient,pan2016analysis}.

  \item[(ii)] The two-level atom is equally coupled ($\kappa_1=\kappa_2=\kappa$) to two identical ($\gamma_1=\gamma_2=\gamma$) incident photons (the blue curve in Fig. \ref{rising}(a)). It is found that the excitation probability attains the maximum value $P_{\rm max}=0.5$ when $\gamma=5\kappa$. Note that in this case, the blue curve in Fig. \ref{rising}(a)  is not symmetric with respect to the time $t=0$ (the part of the curve for $t>0$ coincides with the green curve). However, if we choose $\gamma=\kappa$, then the excitation probability attains its maximum value 0.28 at $t<0$ (the red curve in Fig. \ref{rising}(a)). This reveals the complex dynamics of a two-level atom driven by two counter-propagating photons.

  \item[(iii)] The two incident photons are identical but not equally coupled ($\kappa_1\neq\kappa_2$) to the two-level atom (the purple curve in Fig. \ref{rising}(a)). For instance, the coupling strength between the two-level atom and the photon in the first input channel is $\kappa_1=1$, while the other is $\kappa_2=0.1$. In this case, the interaction between the two-level atom and the photon in the second input channel is relatively weak due to the small value of $\kappa_2$. The maximum excitation probability is $P_{\rm max}=0.77$ when $\gamma_1=\gamma_2=\kappa_1$. Obviously, this excitation probability can be close to $1$ when the coupling $\kappa_2\rightarrow0$ as the limit is the single-photon case (the black curve in Fig. \ref{rising}(a)) as discussed in item (i) above.

  \item[(iv)] The two incident photons are equally coupled ($\kappa_1=\kappa_2=\kappa=1$) to the two-level atom, but with different pulse shapes $\gamma_2=0.01$, $\gamma_1=2\kappa$ (the green curve in Fig. \ref{rising}(a)). In this case, the incident photon in the second channel has a long temporal wave packet (small frequency bandwidth) and is reflected by the atom, see \cite[Fig. 8]{baragiola2012n}.  Moreover, in this parameter setting, the excitation probability reaches its maximum $P_{\rm max}=0.5$.

  \item[(v)] The two incident photons interact with the two-level atom one after another (Fig. \ref{rising}(b)). In this case, the pulse shape of the photon in the first channel is given by
      \begin{equation}
      \xi_1(t)=-\sqrt{\gamma_1}\exp(\frac{\gamma_1}{2}(t+T))H(-T-t),
      \end{equation}
      while the pulse shape of the photon in the second channel is still of the form (\ref{50}). Firstly, we choose $T=10$ (the black curve in Fig. \ref{rising}(b)), that is, the two-level atom can be excited by the incident photon one after another. The excitation probability attains its maximum value $P_{\rm max}=0.5$ when $\gamma=2\kappa$. The same ratio can be found in the \emph{simulated emission} \cite[Fig. 2(a)]{RF16}, where the incident photon interacts with an \emph{excited} atom. In that scenario, the probability of two photons in the second output channel $P_{RR}$ in \cite[Fig. 2(a)]{RF16} attains its maximum when the pulse shape parameter is twice the coupling strength.

  \item[(vi)] It can be observed that the two peak values of the excitation probability get closer as $T$ decreases (the blue curve in Fig. \ref{rising}(b)). Moreover, the excitation probability becomes smaller when the two incident photons are close to each other ($T=1$, the red curve in Fig. \ref{rising}(b)).

  \item[(vii)] Lastly, we set $T=10$, which means the two incident photons are far away from each other. The photon in the second channel is weakly coupled to the two-level atom $\kappa_2=0.1$ (the purple curve in Fig. \ref{rising}(b)). The excitation probability attains its maximum $P_{\rm max}=0.91$ when $\gamma_1=\kappa_1$. However, the second excitation is rather weak due to the weak coupling strength $\kappa_2$.
\end{description}

\subsection{Gaussian pulse shape} \label{sec:Gaussian}

Here, we consider the case that a two-level atom is driven by two counter-propagating photons with the Gaussian pulse shapes
\begin{equation} \label{51}
\xi_i(t)=\left(\frac{\Omega_i^2}{2\pi}\right)^{\frac{1}{4}}\exp\left(-\frac{\Omega_i^2}{4}(t-\tau_i)^2\right),~~i=1,2,
\end{equation}
where $\tau_i$ is the photon peak arrival time and $\Omega_i$ is the frequency bandwidth in the $i$-th input wave packet. If both of the output fields are measured by homodyne detectors, then quantum trajectories are those given in Theorem \ref{theorem}. In this scenario, the excitation probability of the two-level atom  can be calculated by
\begin{equation} \label{Pe_jan9}
P_e(t)=\mathrm{Tr}\left[\rho^{11;11}(t)|e\rangle\langle e|\right],
\end{equation}
where $\rho^{11;11}(t)$ is the solution to (\ref{stochastic}).

\begin{figure}
\centering
\subfigure{\includegraphics[width=0.45\linewidth]{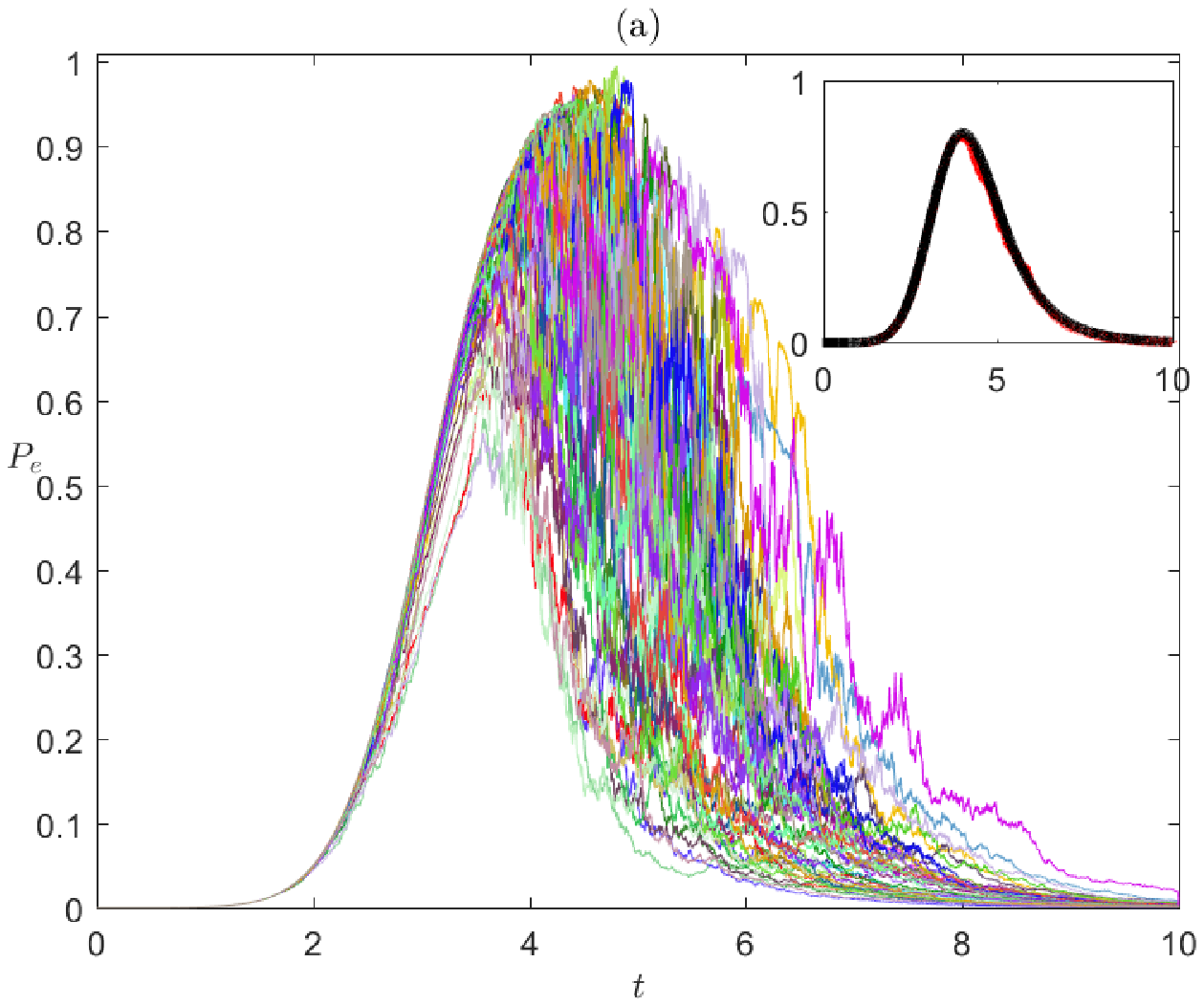}}
\hspace{0.01\linewidth}
\subfigure{\includegraphics[width=0.45\linewidth]{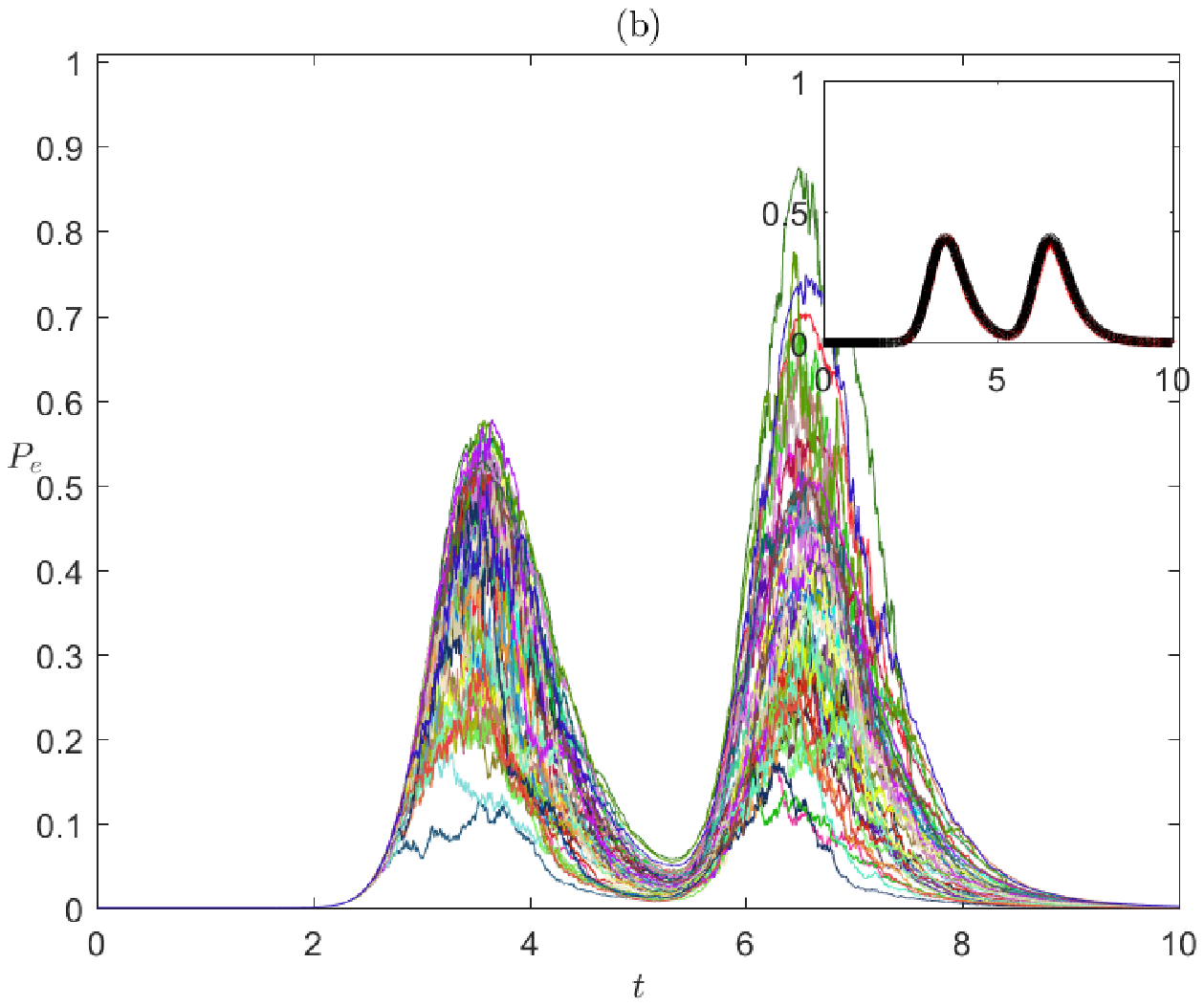}}
\vfill
\subfigure{\includegraphics[width=0.45\linewidth]{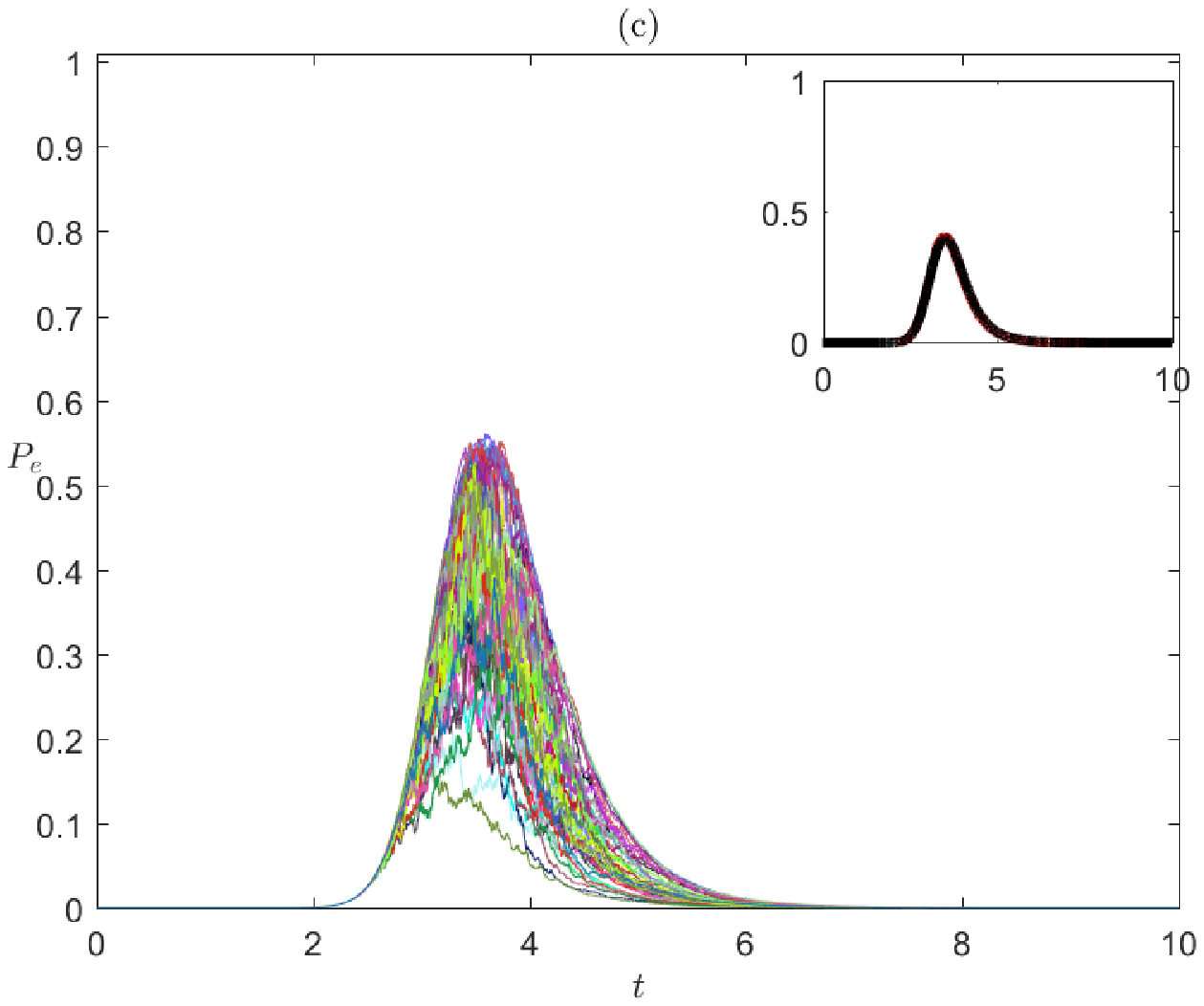}}
\hspace{0.01\linewidth}
\subfigure{\includegraphics[width=0.45\linewidth]{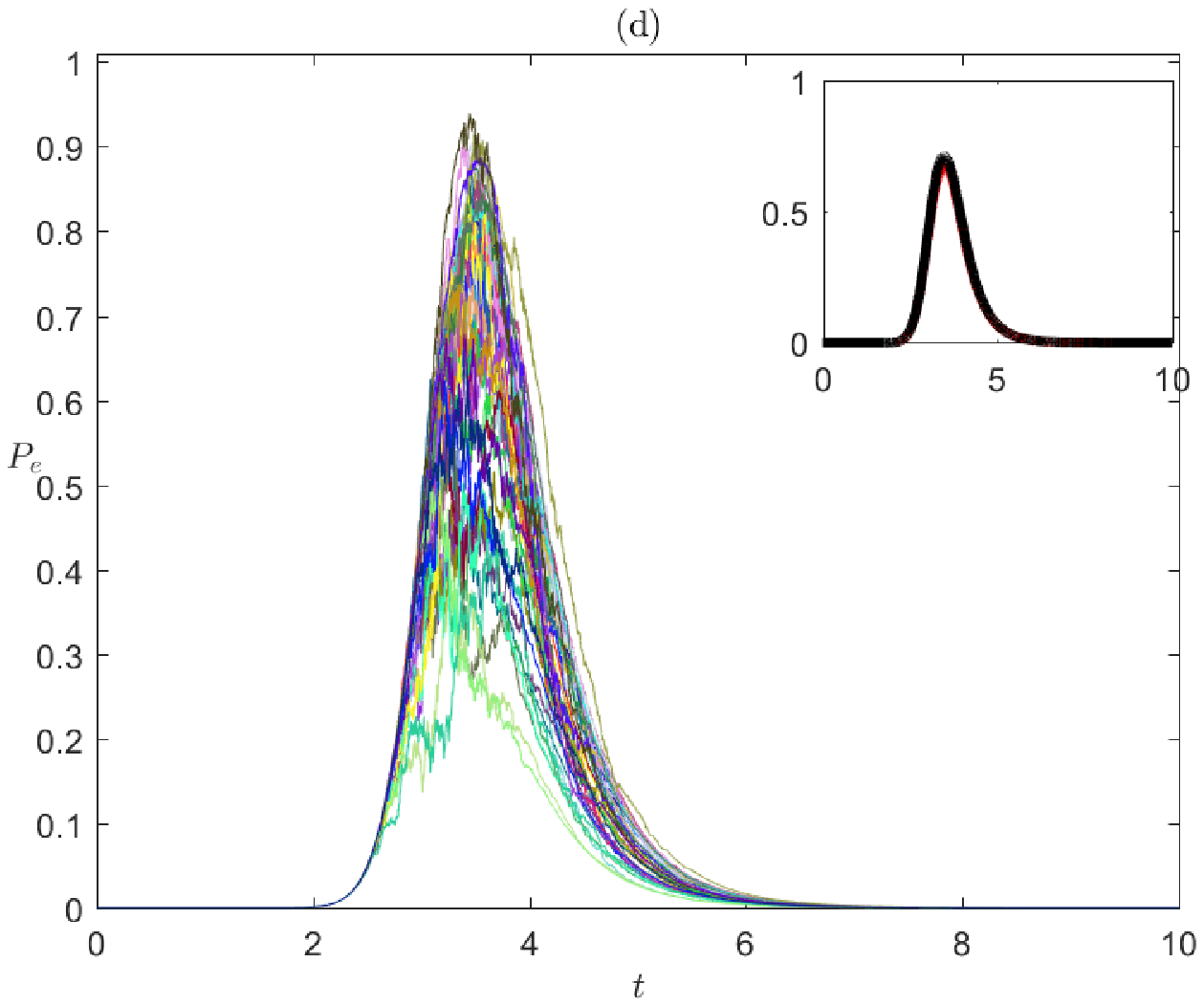}}
\caption{Excitation probabilities of a two-level atom driven by two counter-propagating single-photon states with the Gaussian pulse shapes, calculated by the quantum master equation (the black solid curve in the top-right corner), the quantum trajectories (the fluctuating curves) and their average (the red solid curve  in the top-right corner).}
\label{Gau}
\end{figure}

In what follows, we study the excitation probabilities of the two-level atom by means of the master equations and quantum trajectories. In Fig. \ref{Gau}, the colored fluctuating curves are obtained by means of the quantum filters in Theorem \ref{theorem}, the black solid curves in the top-right corner are calculated by means of the master equation, and the red solid curves in the top-right corner are the results of the average of the quantum trajectories. As can be observed in all subfigures of Fig. \ref{Gau} that the red curves are all very close to the black curves. This confirms the fact that a master equation is an ensemble average of a filtering equation.  We have the following observations.

\begin{description}
  \item[(i)] In Fig. \ref{Gau}(a), we let $\kappa_2=0$. Then we have the single-photon quantum filtering case, See Remark \ref{rem:1channel}. In this case, by the master equation, the maximum of the excitation probability is $0.8$ when the optimal bandwidth $\Omega=1.46\kappa$, see also \cite{SAL09,RSF10,wang2011efficient,GOUGH12QUANTUM,baragiola2012n}.  Moreover, as can be seen in Fig. \ref{Gau}(a) that some individual quantum trajectories can rise up to $P_e=1.0$, which means that the two-level atom may be fully excited. This cannot be observed by the master equation in the top-right corner of Fig. \ref{Gau}(a).

  \item[(ii)] In Fig. \ref{Gau}(b), the two incident photons have different peak arrival times, $\tau_1=3$ and $\tau_2=6$. Based on the master equation it is found that the maximum value of the excitation probability is $P_e=0.4$ when $\Omega_1=2*1.46\kappa_1$ or $\Omega_2=2*1.46\kappa_2$. That is, the two-level atom may be excited twice with the same excitation probability at time instants $t_1=3.5$ and $t_2=6.5$, respectively.  Similar phenomenon can be seen in the scattering of two co-propagating input photons, where the excitation probability is even less than $0.3$ \cite[Fig. 5(a)]{NKMKM15}. Interestingly, the second peak value of some individual quantum trajectories can be significant larger than their first peak value. This cannot be seen by the master equation in the top-right corner of Fig. \ref{Gau}(b).

  \item[(iii)] In Fig. \ref{Gau}(c), the first channel contains a single photon of Gaussian pulse shape, while the second channel is in the vacuum state. By the master equation, the excitation probability $P_e$ is  maximized to $0.4$ when $\Omega_1=2*1.46\kappa_1$. This result is consistent with the Fock-state scattering from a two-level atom \cite[Fig. 7(a)]{baragiola2012n}, in which the maximum value of excitation probability is also $0.4$ when the forward-propagating field is prepared in a Fock state with a single photon. Compared with Fig. \ref{Gau}(a), it can be seen that almost no quantum trajectories can rise beyond $0.6$. In other words, the excitation probability becomes worse due to the vacuum input channel.

  \item[(iv)] In Fig. \ref{Gau}(d), we choose $\tau_1=\tau_2=3$, which means that the two photons interact with the two-level atom simultaneously. By the master equation, the excitation probability can attain its maximum value $P_e=0.71$ at $t=3.5$ when $\Omega_1=2*1.46\kappa_1$ and $\Omega_2=2*1.46\kappa_2$. This optimal bandwidth is consistent with the single-channel two-photon case studied in \cite[Fig. 1]{SONG13MULTI}, where the maximum value of the excitation probability is $0.8796$ when the frequency bandwidths are $\Omega_1=\Omega_2=2*1.46\kappa$. Moreover, there are individual quantum trajectories whose peak values are bigger than 0.9.
\end{description}

\begin{remark}\label{rem:Gaussian}
The interaction between a two-level atom and a single photon or two photons with Gaussian pulse shapes has been studied intensively in the literature. In the single-photon case, when the photon has a Gaussian pulse shape (\ref{51}) with $\Omega=1.46\kappa$, where $\kappa$ is the decay rate of the two-level atom and $\Omega$ is the frequency bandwidth of the single-photon pulse shape, it is shown that the maximal excitation probability is around $0.8$, see, e.g., \cite{SAL09}, \cite{RSF10}, \cite[Fig. 1]{wang2011efficient}, \cite[Fig. 8]{GOUGH12QUANTUM}, \cite[Fig. 2]{baragiola2012n}, and case (i) above. Recently, the analytical expression of the pulse shape of the output single photon has been derived in \cite{pan2016analysis}. Assume the  pulse shape  $\xi(t)$ of the input photon is of the form  (\ref{51}) with photon peak arrival time $\tau=3$ and frequency bandwidth $\Omega=1.46\kappa$. Denote the pulse shape of the output photon by  $\eta(t)$. Then it can be easily verified that $\int_{-\infty}^{{4}}\left(|\xi(\tau)|^2-|\eta(\tau)|^2\right)d\tau=0.8$. Interestingly, the excitation probability achieves its maximum 0.8 at the time $t=4$, see Fig. \ref{Gau}(a). Hence, the filtering result is consistent with the result of input-output response. The two-photon case is more complicated. Quantum filters for a two-level atom driven by a single-channel two-photon state has been derived in \cite{SONG13MULTI}. Numerical simulations show that the maximum value of excitation probability is $0.8796$ when the frequency bandwidths $\Omega_1=\Omega_2=2*1.46\kappa$, see \cite[Fig. 1]{SONG13MULTI}. The analytical form of the output two-photon state has been given in \cite{PDZ16}. Unfortunately, due to the complexity of the output two-photon state, it seems hard to exhibit the consistency between the filtering result and the input-output response result. Interestingly, as shown by the numerical simulations above,  in the case that a two-level atom is driven by two counter-propagating photons with Gaussian pulse shapes, the optimal frequency bandwidths for maximal atomic excitation are  $\Omega_1=2*1.46\kappa_1$ and $\Omega_2=2*1.46\kappa_2$,  which is the same as the single-channel two-photon case studied in \cite{SONG13MULTI}.  These facts indicate that the found scaling (frequency bandwidths are twice of the atomic decay rates) for maximal
atomic excitation may probably have some physical meaning. 
\end{remark}

\section{Conclusion}\label{conclusion}
In this paper, we have investigated the problem of quantum filtering for a two-level atom driven by two  counter-propagating continuous-mode photons. The explicit forms of quantum filters have been derived.  We have used these quantum filters to compute the excitation probabilities when the photons are of rising exponential and Gaussian pulse shapes. Interesting scaling properties between the pulse shape and atomic decay rate have been revealed by these numerical simulations. These simulations illustrate the complex dynamics of the two-level atom driven by two counter-propagating photons, which demands for more rigorous mathematical analysis and experimental exploration.


\begin{acknowledgements}
We wish to thank financial supports from the Hong Kong Research Grant Council under grants 15206915 and 15208418,  JCJC INS2I 2016 ``QIGR3CF'' project and JCJC INS2I 2017 ``QFCCQI'' project.
\end{acknowledgements}

\noindent \textbf{Appendix A.}  In this appendix, we give the stochastic differential equations for the quantum filter mentioned in Theorem \ref{theorem1}. (The dynamics of $\rho^{11;11}(t)$  has already been given in (\ref{HD+Ph}).)

\scriptsize
\begin{eqnarray*}
d\rho^{10;11}(t)&=&\Big\{(\kappa_1+\kappa_2)\mathcal{D}_{\sigma_-}^\star\rho^{10;11}(t)+\sqrt{\kappa_1}\xi_1(t)[\rho^{00;11}(t),\sigma_+]+\sqrt{\kappa_2}\xi_2(t)[\rho^{10;01}(t),\sigma_+]\\
&&+\sqrt{\kappa_2}\xi_2^\ast(t)[\sigma_-,\rho^{10;10}(t)]\Big\}dt\\
&&+\Big\{\sqrt{1-r^2}\left[\xi_1(t)\rho^{00;11}(t)+\sqrt{\kappa_1}\rho^{10;11}(t)\sigma_++\sqrt{\kappa_1}\sigma_-\rho^{10;11}(t)\right]\\
&&+r\left[\xi_2^\ast(t)\rho^{10;10}(t)+\xi_2(t)\rho^{10;01}(t)+\sqrt{\kappa_2}\rho^{10;11}(t)\sigma_++\sqrt{\kappa_2}\sigma_-\rho^{10;11}(t)\right]\\
&&-\rho^{10;11}(t)\left[\sqrt{1-r^2}z_{11}(t)+rz_{12}(t)\right]\Big\}dW_1(t)\\
&&+\bigg\{K_p^{-1}(t)\Big\{r^2\Big[2\sqrt{\kappa_1}\xi_1(t)\rho^{00;11}(t)\sigma_++\kappa_1\sigma_-\rho^{10;11}(t)\sigma_+\Big]\\
&&-r\sqrt{1-r^2}\Big[2\sqrt{\kappa_1}\xi_2(t)\rho^{10;01}(t)\sigma_+\\
&&+2\xi_1(t)\xi_2^\ast(t)\rho^{00;10}(t)+\sqrt{\kappa_1}\xi_2^\ast(t)\sigma_+\rho^{10;10}(t)+2\sqrt{\kappa_2}\xi_1(t)\rho^{00;11}(t)\sigma_+\Big]\\
&&+(1-r^2)\Big[2|\xi_2(t)|^2\rho^{10;00}(t)+\sqrt{\kappa_2}\xi_2^\ast(t)\sigma_-\rho^{10;10}(t)+2\sqrt{\kappa_2}\xi_2(t)\rho^{10;01}(t)\sigma_+\\
&&+\kappa_2\sigma_-\rho^{10;11}(t)\sigma_+\Big]\Big\}-\rho^{10;11}(t)\bigg\}dN(t),\\
d\rho^{00;11}(t)&=&\Big\{(\kappa_1+\kappa_2)\mathcal{D}_{\sigma_-}^\star\rho^{00;11}(t)+\sqrt{\kappa_2}\xi_2(t)[\rho^{00;01}(t),\sigma_+]+\sqrt{\kappa_2}\xi_2^\ast(t)[\sigma_-,\rho^{00;10}(t)]\Big\}dt\\
&&+\Big\{\sqrt{1-r^2}\left[\sqrt{\kappa_1}\rho^{00;11}(t)\sigma_++\sqrt{\kappa_1}\sigma_-\rho^{00;11}(t)\right]\\
&&+r\left[\xi_2^\ast(t)\rho^{00;10}(t)+\xi_2(t)\rho^{00;01}(t)+\sqrt{\kappa_2}\rho^{00;11}(t)\sigma_++\sqrt{\kappa_2}\sigma_-\rho^{00;11}(t)\right]\\
&&-\rho^{00;11}(t)\left[\sqrt{1-r^2}z_{11}(t)+rz_{12}(t)\right]\Big\}dW_1(t)\\
&&+\bigg\{K_p^{-1}(t)\Big\{r^2\Big[\kappa_1\sigma_-\rho^{00;11}(t)\sigma_+\Big]-r\sqrt{1-r^2}\Big[2\sqrt{\kappa_1}\xi_2(t)\rho^{00;01}(t)\sigma_+\\
&&+\sqrt{\kappa_1}\xi_2^\ast(t)\sigma_-\rho^{00;10}(t)\Big]+(1-r^2)\Big[2|\xi_2(t)|^2\rho^{00;00}(t)+\sqrt{\kappa_2}\xi_2^\ast(t)\sigma_-\rho^{00;10}(t)\\
&&+2\sqrt{\kappa_2}\xi_2(t)\rho^{00;01}(t)\sigma_++\kappa_2\sigma_-\rho^{00;11}(t)\sigma_+\Big]\Big\}-\rho^{00;11}(t)\bigg\}dN(t),\\
d\rho^{11;10}(t)&=&\Big\{(\kappa_1+\kappa_2)\mathcal{D}_{\sigma_-}^\star\rho^{11;10}(t)+\sqrt{\kappa_1}\xi_1(t)[\rho^{01;10}(t),\sigma_+]+\sqrt{\kappa_1}\xi_1^\ast(t)[\sigma_-,\rho^{10;10}(t)]\\
&&+\sqrt{\kappa_2}\xi_2(t)[\rho^{11;00}(t),\sigma_+]\Big\}dt\\
&&+\Big\{\sqrt{1-r^2}\left[\xi_1^\ast(t)\rho^{10;10}(t)+\xi_1(t)\rho^{01;10}(t)+\sqrt{\kappa_1}\rho^{11;10}(t)\sigma_++\sqrt{\kappa_1}\sigma_-\rho^{11;10}(t)\right]\\
&&+r\left[\xi_2(t)\rho^{11;00}(t)+\sqrt{\kappa_2}\rho^{11;10}(t)\sigma_++\sqrt{\kappa_2}\sigma_-\rho^{11;10}(t)\right]\\
&&-\rho^{11;10}(t)\left[\sqrt{1-r^2}z_{11}(t)+rz_{12}(t)\right]\Big\}dW_1(t)\\
&&+\bigg\{K_p^{-1}(t)\Big\{r^2\Big[2|\xi_1(t)|^2\rho^{00;10}(t)+\sqrt{\kappa_1}\xi_1^\ast(t)\sigma_-\rho^{10;10}(t)+2\sqrt{\kappa_1}\xi_1(t)\rho^{01;10}(t)\sigma_+\\
&&+\kappa_1\sigma_-\rho^{11;10}(t)\sigma_+\Big]\\
&&-r\sqrt{1-r^2}\Big[2\xi_1^\ast(t)\xi_2(t)\rho^{10;00}(t)+\sqrt{\kappa_2}\xi_1^\ast(t)\sigma_-\rho^{10;10}(t)+2\sqrt{\kappa_1}\xi_2(t)\rho^{11;00}(t)\sigma_+\\
&&+2\sqrt{\kappa_2}\xi_1(t)\rho^{01;10}(t)\sigma_+\Big]+(1-r^2)\Big[2\sqrt{\kappa_2}\xi_2(t)\rho^{11;00}(t)\sigma_++\kappa_2\sigma_-\rho^{11;10}(t)\sigma_+\Big]\Big\}\\
&&-\rho^{11;10}(t)\bigg\}dN(t),\\
d\rho^{10;10}(t)&=&\Big\{(\kappa_1+\kappa_2)\mathcal{D}_{\sigma_-}^\star\rho^{10;10}(t)+\sqrt{\kappa_1}\xi_1(t)[\rho^{00;10}(t),\sigma_+]+\sqrt{\kappa_2}\xi_2(t)[\rho^{10;00}(t),\sigma_+]\Big\}dt\\
&&+\Big\{\sqrt{1-r^2}\left[\xi_1(t)\rho^{00;10}(t)+\sqrt{\kappa_1}\rho^{10;10}(t)\sigma_++\sqrt{\kappa_1}\sigma_-\rho^{10;10}(t)\right]\\
&&+r\left[\xi_2(t)\rho^{10;00}(t)+\sqrt{\kappa_2}\rho^{10;10}(t)\sigma_++\sqrt{\kappa_2}\sigma_-\rho^{10;10}(t)\right]\\
&&-\rho^{10;10}(t)\left[\sqrt{1-r^2}z_{11}(t)+rz_{12}(t)\right]\Big\}dW_1(t)\\
&&+\bigg\{K_p^{-1}(t)\Big\{r^2\Big[2\sqrt{\kappa_1}\xi_1(t)\rho^{00;10}(t)\sigma_++\kappa_1\sigma_-\rho^{10;10}(t)\sigma_+\Big]\\
&&-r\sqrt{1-r^2}\Big[2\sqrt{\kappa_1}\xi_2(t)\rho^{10;00}(t)\sigma_++2\sqrt{\kappa_2}\xi_1(t)\rho^{00;10}(t)\sigma_+\Big]\\
&&+(1-r^2)\Big[2\sqrt{\kappa_2}\xi_2(t)\rho^{10;00}(t)\sigma_++\kappa_2\sigma_-\rho^{10;10}(t)\sigma_+\Big]\Big\}-\rho^{10;10}(t)\bigg\}dN(t),\\
d\rho^{01;10}(t)&=&\Big\{(\kappa_1+\kappa_2)\mathcal{D}_{\sigma_-}^\star\rho^{01;10}(t)+\sqrt{\kappa_1}\xi_1^\ast(t)[\sigma_-,\rho^{00;10}(t)]+\sqrt{\kappa_2}\xi_2(t)[\rho^{01;00}(t),\sigma_+]\Big\}dt\\
&&+\Big\{\sqrt{1-r^2}\left[\xi_1^\ast(t)\rho^{00;10}(t)+\sqrt{\kappa_1}\rho^{01;10}(t)\sigma_++\sqrt{\kappa_1}\sigma_-\rho^{01;10}(t)\right]\\
&&+r\left[\xi_2(t)\rho^{01;00}(t)+\sqrt{\kappa_2}\rho^{01;10}(t)\sigma_++\sqrt{\kappa_2}\sigma_-\rho^{01;10}(t)\right]\\
&&-\rho^{01;10}(t)\left[\sqrt{1-r^2}z_{11}(t)+rz_{12}(t)\right]\Big\}dW_1(t)\\
&&+\bigg\{K_p^{-1}(t)\Big\{r^2\Big[\sqrt{\kappa_1}\xi_1^\ast(t)\sigma_-\rho^{00;10}(t)+\kappa_1\sigma_-\rho^{01;10}(t)\sigma_+\Big]\\
&&-r\sqrt{1-r^2}\Big[2\xi_1^\ast(t)\xi_2(t)\rho^{00;00}(t)+\sqrt{\kappa_2}\xi_1^\ast(t)\sigma_-\rho^{00;10}(t)+2\sqrt{\kappa_1}\xi_2(t)\rho^{01;00}(t)\sigma_+\Big]\\
&&+(1-r^2)\Big[2\sqrt{\kappa_2}\xi_2(t)\rho^{01;00}(t)\sigma_++\kappa_2\sigma_-\rho^{01;10}(t)\sigma_+\Big]\Big\}-\rho^{01;10}(t)\bigg\}dN(t),\\
d\rho^{00;10}(t)&=&\Big\{(\kappa_1+\kappa_2)\mathcal{D}_{\sigma_-}^\star\rho^{00;10}(t)+\sqrt{\kappa_2}\xi_2(t)[\rho^{00;00}(t),\sigma_+]\Big\}dt\\
&&+\Big\{\sqrt{1-r^2}\left[\sqrt{\kappa_1}\rho^{00;10}(t)\sigma_++\sqrt{\kappa_1}\sigma_-\rho^{00;10}(t)\right]+r\big[\xi_2(t)\rho^{00;00}(t)\\
&&+\sqrt{\kappa_2}\rho^{00;10}(t)\sigma_++\sqrt{\kappa_2}\sigma_-\rho^{00;10}(t)\big]-\rho^{00;10}(t)\left[\sqrt{1-r^2}z_{11}(t)+rz_{12}(t)\right]\Big\}dW_1(t)\\
&&+\bigg\{K_p^{-1}(t)\Big\{r^2\Big[\kappa_1\sigma_-\rho^{00;10}(t)\sigma_+\Big]-r\sqrt{1-r^2}\Big[2\sqrt{\kappa_1}\xi_2(t)\rho^{00;00}(t)\sigma_+\Big]\\
&&+(1-r^2)\Big[2\sqrt{\kappa_2}\xi_2(t)\rho^{00;00}(t)\sigma_++\kappa_2\sigma_-\rho^{00;10}(t)\sigma_+\Big]\Big\}-\rho^{00;10}(t)\bigg\}dN(t),\\
d\rho^{11;00}(t)&=&\Big\{(\kappa_1+\kappa_2)\mathcal{D}_{\sigma_-}^\star\rho^{11;00}(t)+\sqrt{\kappa_1}\xi_1(t)[\rho^{01;00}(t),\sigma_+]+\sqrt{\kappa_1}\xi_1^\ast(t)[\sigma_-,\rho^{10;00}(t)]\Big\}dt\\
&&+\Big\{\sqrt{1-r^2}\left[\xi_1^\ast(t)\rho^{10;00}(t)+\xi_1(t)\rho^{01;00}(t)+\sqrt{\kappa_1}\rho^{11;00}(t)\sigma_++\sqrt{\kappa_1}\sigma_-\rho^{11;00}(t)\right]\\
&&+r\left[\sqrt{\kappa_2}\rho^{11;00}(t)\sigma_++\sqrt{\kappa_2}\sigma_-\rho^{11;00}(t)\right]-\rho^{11;00}(t)\left[\sqrt{1-r^2}z_{11}(t)+rz_{12}(t)\right]\Big\}dW_1(t)\\
&&+\bigg\{K_p^{-1}(t)\Big\{r^2\Big[2|\xi_1(t)|^2\rho^{00;00}(t)+\sqrt{\kappa_1}\xi_1^\ast(t)\sigma_-\rho^{10;00}(t)\\
&&+2\sqrt{\kappa_1}\xi_1(t)\rho^{01;00}(t)\sigma_++\kappa_1\sigma_-\rho^{11;00}(t)\sigma_+\Big]\\
&&-r\sqrt{1-r^2}\Big[\sqrt{\kappa_2}\xi_1^\ast(t)\sigma_-\rho^{10;00}(t)+2\sqrt{\kappa_2}\xi_1(t)\rho^{01;00}(t)\sigma_+\Big]\\
&&+(1-r^2)\Big[\kappa_2\sigma_-\rho^{11;00}(t)\sigma_+\Big]\Big\}-\rho^{11;00}(t)\bigg\}dN(t),\\
d\rho^{10;00}(t)&=&\Big\{(\kappa_1+\kappa_2)\mathcal{D}_{\sigma_-}^\star\rho^{10;00}(t)+\sqrt{\kappa_1}\xi_1(t)[\rho^{00;00}(t),\sigma_+]\Big\}dt\\
&&+\Big\{\sqrt{1-r^2}\left[\xi_1(t)\rho^{00;00}(t)+\sqrt{\kappa_1}\rho^{10;00}(t)\sigma_++\sqrt{\kappa_1}\sigma_-\rho^{10;00}(t)\right]\\
&&+r\left[\sqrt{\kappa_2}\rho^{10;00}(t)\sigma_++\sqrt{\kappa_2}\sigma_-\rho^{10;00}(t)\right]\\
&&-\rho^{10;00}(t)\left[\sqrt{1-r^2}z_{11}(t)+rz_{12}(t)\right]\Big\}dW_1(t)\\
&&+\bigg\{K_p^{-1}(t)\Big\{r^2\Big[2\sqrt{\kappa_1}\xi_1(t)\rho^{00;00}(t)\sigma_++\kappa_1\sigma_-\rho^{10;00}(t)\sigma_+\Big]\\
&&-r\sqrt{1-r^2}\Big[2\sqrt{\kappa_2}\xi_1(t)\rho^{00;00}(t)\sigma_+\Big]\\
&&+(1-r^2)\Big[\kappa_2\sigma_-\rho^{10;00}(t)\sigma_+\Big]\Big\}-\rho^{10;00}(t)\bigg\}dN(t),\\
d\rho^{00;00}(t)&=&\Big\{(\kappa_1+\kappa_2)\mathcal{D}_{\sigma_-}^\star\rho^{00;00}(t)\Big\}dt\\
&&+\Big\{\sqrt{1-r^2}\left[\sqrt{\kappa_1}\rho^{00;00}(t)\sigma_++\sqrt{\kappa_1}\sigma_-\rho^{00;00}(t)\right]+r\big[\sqrt{\kappa_2}\rho^{00;00}(t)\sigma_+\\
&&+\sqrt{\kappa_2}\sigma_-\rho^{00;00}(t)\big]-\rho^{00;00}(t)\left[\sqrt{1-r^2}z_{11}(t)+rz_{12}(t)\right]\Big\}dW_1(t)\\
&&+\bigg\{K_p^{-1}(t)\Big\{r^2\Big[\kappa_1\sigma_-\rho^{00;00}(t)\sigma_+\Big]+(1-r^2)\Big[\kappa_2\sigma_-\rho^{00;00}(t)\sigma_+\Big]\Big\}\\
&&-\rho^{00;00}(t)\bigg\}dN(t).
\end{eqnarray*}
\normalsize



\noindent \textbf{Appendix B.}  In this appendix, we give the stochastic differential equations for the quantum filter mentioned in Theorem \ref{theorem}. (The dynamics of $\rho^{11;11}(t)$ has already been given in (\ref{stochastic}).)

\scriptsize
\begin{eqnarray*}
d\rho^{10;11}(t)&=&\Big\{(\kappa_1+\kappa_2)\mathcal{D}_{\sigma_-}^\star\rho^{10;11}(t)+\sqrt{\kappa_1}\xi_1(t)[\rho^{00;11}(t),\sigma_+]+\sqrt{\kappa_2}\xi_2(t)[\rho^{10;01}(t),\sigma_+]\\
&&+\sqrt{\kappa_2}\xi_2^\ast(t)[\sigma_-,\rho^{10;10}(t)]\Big\}dt\\
&&+\Big\{\sqrt{1-r^2}\left[\xi_1(t)\rho^{00;11}(t)+\sqrt{\kappa_1}\rho^{10;11}(t)\sigma_++\sqrt{\kappa_1}\sigma_-\rho^{10;11}(t)\right]\\
&&+r\left[\xi_2^\ast(t)\rho^{10;10}(t)+\xi_2(t)\rho^{10;01}(t)+\sqrt{\kappa_2}\rho^{10;11}(t)\sigma_++\sqrt{\kappa_2}\sigma_-\rho^{10;11}(t)\right]\\
&&-\rho^{10;11}(t)\left[\sqrt{1-r^2}z_{11}(t)+rz_{12}(t)\right]\Big\}dW_1(t)\\
&&+\Big\{-r\left[\xi_1(t)\rho^{00;11}(t)+\sqrt{\kappa_1}\rho^{10;11}(t)\sigma_++\sqrt{\kappa_1}\sigma_-\rho^{10;11}(t)\right]\\
&&+\sqrt{1-r^2}\left[\xi_2^\ast(t)\rho^{10;10}(t)+\xi_2(t)\rho^{10;01}(t)+\sqrt{\kappa_2}\rho^{10;11}(t)\sigma_++\sqrt{\kappa_2}\sigma_-\rho^{10;11}(t)\right]\\
&&-\rho^{10;11}(t)\left[-rz_{11}(t)+\sqrt{1-r^2}z_{12}(t)\right]\Big\}dW_2(t),\\
d\rho^{00;11}(t)&=&\Big\{(\kappa_1+\kappa_2)\mathcal{D}_{\sigma_-}^\star\rho^{00;11}(t)+\sqrt{\kappa_2}\xi_2(t)[\rho^{00;01}(t),\sigma_+]+\sqrt{\kappa_2}\xi_2^\ast(t)[\sigma_-,\rho^{00;10}(t)]\Big\}dt\\
&&+\Big\{\sqrt{1-r^2}\left[\sqrt{\kappa_1}\rho^{00;11}(t)\sigma_++\sqrt{\kappa_1}\sigma_-\rho^{00;11}(t)\right]\\
&&+r\left[\xi_2^\ast(t)\rho^{00;10}(t)+\xi_2(t)\rho^{00;01}(t)+\sqrt{\kappa_2}\rho^{00;11}(t)\sigma_++\sqrt{\kappa_2}\sigma_-\rho^{00;11}(t)\right]\\
&&-\rho^{00;11}(t)\left[\sqrt{1-r^2}z_{11}(t)+rz_{12}(t)\right]\Big\}dW_1(t)\\
&&+\Big\{-r\left[\sqrt{\kappa_1}\rho^{00;11}(t)\sigma_++\sqrt{\kappa_1}\sigma_-\rho^{00;11}(t)\right]\\
&&+\sqrt{1-r^2}\left[\xi_2^\ast(t)\rho^{00;10}(t)+\xi_2(t)\rho^{00;01}(t)+\sqrt{\kappa_2}\rho^{00;11}(t)\sigma_++\sqrt{\kappa_2}\sigma_-\rho^{00;11}(t)\right]\\
&&-\rho^{00;11}(t)\left[-rz_{11}(t)+\sqrt{1-r^2}z_{12}(t)\right]\Big\}dW_2(t),\\
d\rho^{11;10}(t)&=&\Big\{(\kappa_1+\kappa_2)\mathcal{D}_{\sigma_-}^\star\rho^{11;10}(t)+\sqrt{\kappa_1}\xi_1(t)[\rho^{01;10}(t),\sigma_+]+\sqrt{\kappa_1}\xi_1^\ast(t)[\sigma_-,\rho^{10;10}(t)]\\
&&+\sqrt{\kappa_2}\xi_2(t)[\rho^{11;00}(t),\sigma_+]\Big\}dt\\
&&+\Big\{\sqrt{1-r^2}\left[\xi_1^\ast(t)\rho^{10;10}(t)+\xi_1(t)\rho^{01;10}(t)+\sqrt{\kappa_1}\rho^{11;10}(t)\sigma_++\sqrt{\kappa_1}\sigma_-\rho^{11;10}(t)\right]\\
&&+r\left[\xi_2(t)\rho^{11;00}(t)+\sqrt{\kappa_2}\rho^{11;10}(t)\sigma_++\sqrt{\kappa_2}\sigma_-\rho^{11;10}(t)\right]\\
&&-\rho^{11;10}(t)\left[\sqrt{1-r^2}z_{11}(t)+rz_{12}(t)\right]\Big\}dW_1(t)\\
&&+\Big\{-r\left[\xi_1^\ast(t)\rho^{10;10}(t)+\xi_1(t)\rho^{01;10}(t)+\sqrt{\kappa_1}\rho^{11;10}(t)\sigma_++\sqrt{\kappa_1}\sigma_-\rho^{11;10}(t)\right]\\
&&+\sqrt{1-r^2}\left[\xi_2(t)\rho^{11;00}(t)+\sqrt{\kappa_2}\rho^{11;10}(t)\sigma_++\sqrt{\kappa_2}\sigma_-\rho^{11;10}(t)\right]\\
&&-\rho^{11;10}(t)\left[-rz_{11}(t)+\sqrt{1-r^2}z_{12}(t)\right]\Big\}dW_2(t),\\
d\rho^{10;10}(t)&=&\Big\{(\kappa_1+\kappa_2)\mathcal{D}_{\sigma_-}^\star\rho^{10;10}(t)+\sqrt{\kappa_1}\xi_1(t)[\rho^{00;10}(t),\sigma_+]+\sqrt{\kappa_2}\xi_2(t)[\rho^{10;00}(t),\sigma_+]\Big\}dt\\
&&+\Big\{\sqrt{1-r^2}\left[\xi_1(t)\rho^{00;10}(t)+\sqrt{\kappa_1}\rho^{10;10}(t)\sigma_++\sqrt{\kappa_1}\sigma_-\rho^{10;10}(t)\right]\\
&&+r\left[\xi_2(t)\rho^{10;00}(t)+\sqrt{\kappa_2}\rho^{10;10}(t)\sigma_++\sqrt{\kappa_2}\sigma_-\rho^{10;10}(t)\right]\\
&&-\rho^{10;10}(t)\left[\sqrt{1-r^2}z_{11}(t)+rz_{12}(t)\right]\Big\}dW_1(t)\\
&&+\Big\{-r\left[\xi_1(t)\rho^{00;10}(t)+\sqrt{\kappa_1}\rho^{10;10}(t)\sigma_++\sqrt{\kappa_1}\sigma_-\rho^{10;10}(t)\right]\\
&&+\sqrt{1-r^2}\left[\xi_2(t)\rho^{10;00}(t)+\sqrt{\kappa_2}\rho^{10;10}(t)\sigma_++\sqrt{\kappa_2}\sigma_-\rho^{10;10}(t)\right]\\
&&-\rho^{10;10}(t)\left[-rz_{11}(t)+\sqrt{1-r^2}z_{12}(t)\right]\Big\}dW_2(t),\\
d\rho^{01;10}(t)&=&\Big\{(\kappa_1+\kappa_2)\mathcal{D}_{\sigma_-}^\star\rho^{01;10}(t)+\sqrt{\kappa_1}\xi_1^\ast(t)[\sigma_-,\rho^{00;10}(t)]+\sqrt{\kappa_2}\xi_2(t)[\rho^{01;00}(t),\sigma_+]\Big\}dt\\
&&+\Big\{\sqrt{1-r^2}\left[\xi_1^\ast(t)\rho^{00;10}(t)+\sqrt{\kappa_1}\rho^{01;10}(t)\sigma_++\sqrt{\kappa_1}\sigma_-\rho^{01;10}(t)\right]\\
&&+r\left[\xi_2(t)\rho^{01;00}(t)+\sqrt{\kappa_2}\rho^{01;10}(t)\sigma_++\sqrt{\kappa_2}\sigma_-\rho^{01;10}(t)\right]\\
&&-\rho^{01;10}(t)\left[\sqrt{1-r^2}z_{11}(t)+rz_{12}(t)\right]\Big\}dW_1(t)\\
&&+\Big\{-r\left[\xi_1^\ast(t)\rho^{00;10}(t)+\sqrt{\kappa_1}\rho^{01;10}(t)\sigma_++\sqrt{\kappa_1}\sigma_-\rho^{01;10}(t)\right]\\
&&+\sqrt{1-r^2}\left[\xi_2(t)\rho^{01;00}(t)+\sqrt{\kappa_2}\rho^{01;10}(t)\sigma_++\sqrt{\kappa_2}\sigma_-\rho^{01;10}(t)\right]\\
&&-\rho^{01;10}(t)\left[-rz_{11}(t)+\sqrt{1-r^2}z_{12}(t)\right]\Big\}dW_2(t),\\
d\rho^{00;10}(t)&=&\Big\{(\kappa_1+\kappa_2)\mathcal{D}_{\sigma_-}^\star\rho^{00;10}(t)+\sqrt{\kappa_2}\xi_2(t)[\rho^{00;00}(t),\sigma_+]\Big\}dt\\
&&+\Big\{\sqrt{1-r^2}\left[\sqrt{\kappa_1}\rho^{00;10}(t)\sigma_++\sqrt{\kappa_1}\sigma_-\rho^{00;10}(t)\right]+r\big[\xi_2(t)\rho^{00;00}(t)\\
&&+\sqrt{\kappa_2}\rho^{00;10}(t)\sigma_++\sqrt{\kappa_2}\sigma_-\rho^{00;10}(t)\big]-\rho^{00;10}(t)\left[\sqrt{1-r^2}z_{11}(t)+rz_{12}(t)\right]\Big\}dW_1(t)\\
&&+\Big\{-r\left[\sqrt{\kappa_1}\rho^{00;10}(t)\sigma_++\sqrt{\kappa_1}\sigma_-\rho^{00;10}(t)\right]+\sqrt{1-r^2}\big[\xi_2(t)\rho^{00;00}(t)\\
&&+\sqrt{\kappa_2}\rho^{00;10}(t)\sigma_++\sqrt{\kappa_2}\sigma_-\rho^{00;10}(t)\big]-\rho^{00;10}(t)\left[-rz_{11}(t)+\sqrt{1-r^2}z_{12}(t)\right]\Big\}dW_2(t),\\
d\rho^{11;00}(t)&=&\Big\{(\kappa_1+\kappa_2)\mathcal{D}_{\sigma_-}^\star\rho^{11;00}(t)+\sqrt{\kappa_1}\xi_1(t)[\rho^{01;00}(t),\sigma_+]+\sqrt{\kappa_1}\xi_1^\ast(t)[\sigma_-,\rho^{10;00}(t)]\Big\}dt\\
&&+\Big\{\sqrt{1-r^2}\left[\xi_1^\ast(t)\rho^{10;00}(t)+\xi_1(t)\rho^{01;00}(t)+\sqrt{\kappa_1}\rho^{11;00}(t)\sigma_++\sqrt{\kappa_1}\sigma_-\rho^{11;00}(t)\right]\\
&&+r\left[\sqrt{\kappa_2}\rho^{11;00}(t)\sigma_++\sqrt{\kappa_2}\sigma_-\rho^{11;00}(t)\right]-\rho^{11;00}(t)\left[\sqrt{1-r^2}z_{11}(t)+rz_{12}(t)\right]\Big\}dW_1(t)\\
&&+\Big\{-r\left[\xi_1^\ast(t)\rho^{10;00}(t)+\xi_1(t)\rho^{01;00}(t)+\sqrt{\kappa_1}\rho^{11;00}(t)\sigma_++\sqrt{\kappa_1}\sigma_-\rho^{11;00}(t)\right]\\
&&+\sqrt{1-r^2}\left[\sqrt{\kappa_2}\rho^{11;00}(t)\sigma_++\sqrt{\kappa_2}\sigma_-\rho^{11;00}(t)\right]\\
&&-\rho^{11;00}(t)\left[-rz_{11}(t)+\sqrt{1-r^2}z_{12}(t)\right]\Big\}dW_2(t),\\
d\rho^{10;00}(t)&=&\Big\{(\kappa_1+\kappa_2)\mathcal{D}_{\sigma_-}^\star\rho^{10;00}(t)+\sqrt{\kappa_1}\xi_1(t)[\rho^{00;00}(t),\sigma_+]\Big\}dt\\
&&+\Big\{\sqrt{1-r^2}\left[\xi_1(t)\rho^{00;00}(t)+\sqrt{\kappa_1}\rho^{10;00}(t)\sigma_++\sqrt{\kappa_1}\sigma_-\rho^{10;00}(t)\right]\\
&&+r\left[\sqrt{\kappa_2}\rho^{10;00}(t)\sigma_++\sqrt{\kappa_2}\sigma_-\rho^{10;00}(t)\right]\\
&&-\rho^{10;00}(t)\left[\sqrt{1-r^2}z_{11}(t)+rz_{12}(t)\right]\Big\}dW_1(t)\\
&&+\Big\{-r\left[\xi_1(t)\rho^{00;00}(t)+\sqrt{\kappa_1}\rho^{10;00}(t)\sigma_++\sqrt{\kappa_1}\sigma_-\rho^{10;00}(t)\right]\\
&&+\sqrt{1-r^2}\left[\sqrt{\kappa_2}\rho^{10;00}(t)\sigma_++\sqrt{\kappa_2}\sigma_-\rho^{10;00}(t)\right]\\
&&-\rho^{10;00}(t)\left[-rz_{11}(t)+\sqrt{1-r^2}z_{12}(t)\right]\Big\}dW_2(t),\\
d\rho^{00;00}(t)&=&\Big\{(\kappa_1+\kappa_2)\mathcal{D}_{\sigma_-}^\star\rho^{00;00}(t)\Big\}dt\\
&&+\Big\{\sqrt{1-r^2}\left[\sqrt{\kappa_1}\rho^{00;00}(t)\sigma_++\sqrt{\kappa_1}\sigma_-\rho^{00;00}(t)\right]+r\big[\sqrt{\kappa_2}\rho^{00;00}(t)\sigma_+\\
&&+\sqrt{\kappa_2}\sigma_-\rho^{00;00}(t)\big]-\rho^{00;00}(t)\left[\sqrt{1-r^2}z_{11}(t)+rz_{12}(t)\right]\Big\}dW_1(t)\\
&&+\Big\{-r\left[\sqrt{\kappa_1}\rho^{00;00}(t)\sigma_++\sqrt{\kappa_1}\sigma_-\rho^{00;00}(t)\right]+\sqrt{1-r^2}\big[\sqrt{\kappa_2}\rho^{00;00}(t)\sigma_+\\
&&+\sqrt{\kappa_2}\sigma_-\rho^{00;00}(t)\big]-\rho^{00;00}(t)\left[-rz_{11}(t)+\sqrt{1-r^2}z_{12}(t)\right]\Big\}dW_2(t).
\end{eqnarray*}
\normalsize


\end{document}